\documentclass[usenatbib,usegraphicx]{mn2e}
\pdfoutput=1

\usepackage{color}
\usepackage{times}

\def\solar{\ifmmode_{\odot}\else$_{\odot}$\fi}

\newcommand{\hcop}{HCO$^+$}

\def\kms{\ifmmode{\rm \,km\,s}^{-1}\else\,km\,s$^{-1}$\fi}
\newcommand{\degree}{$^{\circ}$}
\newcommand{\fdeg}{$^{\circ}$\hspace{-1mm}.}

\def\lapp{\ifmmode\stackrel{<}{_{\sim}}\else$\stackrel{<}{_{\sim}}$\fi}
\def\gapp{\ifmmode\stackrel{>}{_{\sim}}\else$\stackrel{>}{_{\sim}}$\fi}
\newcommand{\cc}{CanariCam}

\title[Magnetic Feld Structures in K3-50]{Magnetic Field Structures in Star Forming Regions: 
    Mid-Infrared Imaging Polarimetry of K3-50}
\author[Barnes et al.]{Peter Barnes$^{1,2}$\thanks{E-mail: pjb@astro.ufl.edu}, Dan Li$^{1}$, Charles Telesco$^{1}$, Nahathai Tanakul$^{1}$, Naib\'i Mari\~nas$^{1}$, 
 \newauthor
Chris Wright$^{3}$, Chris Packham$^{4}$, Eric Pantin$^{1,5}$, Patrick Roche$^{6}$, James Hough$^{7}$\vspace{2mm}\\ 
$^{1}$Astronomy Department, University of Florida, Gainesville, FL 32611, USA\\
$^{2}$School of Science and Technology, University of New England, Armidale NSW 2351, Australia\\
$^{3}$School of Physical, Environmental, and Mathematical Sciences, University of New South Wales, Canberra, ACT 2610, Australia\\
$^{4}$Physics and Astronomy Department, University of Texas at San Antonio, 1 UTSA Circle, San Antonio, TX 78249\\
$^{5}$Laboratoire AIM, CEA/DSM - CNRS - Universit\'e Paris Diderot, IRFU/SAp, 91191 Gif sur Yvette, France\\
$^{6}$Astrophysics, Department of Physics, University of Oxford, Keble Road, Oxford OX1 3RH, UK\\
$^{7}$Centre for Astrophysics Research, University of Hertfordshire, Hatfield, AL10 9AB, UK
}

\begin{document}

\setcounter{page}{1}
\date{Accepted 2015 June 4. Received 2015 June 3; in original form 2015 March 24}

\pagerange{\pageref{firstpage}--\pageref{lastpage}} \pubyear{2015}

\maketitle

\label{firstpage}

\begin{abstract}
We report new imaging polarimetry observations of the Galactic compact HII region K3-50 using \cc\ at the Gran Telescopio Canarias.  We use a standard polarimetric analysis technique, first outlined by Aitken, to decompose the observed polarisation images centred at 8.7, 10.3, and 12.5 $\mu$m into the emissive and absorptive components from silicate grains that are aligned with the local magnetic field.  These components reveal the spatially-resolved magnetic field structures across the mid-infrared emission area of K3-50.  We examine these structures and show that they are consistent with previously observed features and physical models of K3-50, such as the molecular torus and the ionised outflow.  We propose a 3D geometry for all the structures seen at different wavelengths.  We also compute relevant physical quantities in order to estimate the associated magnetic field strengths that would be implied under various physical assumptions.  We compare these results with MHD simulations of protostar formation that predict the magnetic field strength and configuration.  We find that the magnetic field may be dynamically important in the innermost 0.2\,pc of the molecular torus, but that the torus is more likely to be rotationally-supported against gravity outside this radius.  Similarly, magnetic fields are unlikely to dominate the {\em global} physics of the ionised outflow, but they may be important in helping confine the flow near the cavity wall in some locations.  Ours is the first application of the Aitken technique to spatially-resolved magnetic field structures in multiple layers along the line of sight, effectively a method of ``polarisation tomography.''  
\end{abstract}

\begin{keywords}
stars: circumstellar matter --- techniques: polarimetric --- ISM: magnetic fields --- ISM: individual 
(K3-50) --- stars: formation.
\end{keywords}

\section{Introduction}
Magnetic fields may play a number of important roles in various astrophysical settings, such as in AGN (synchrotron emission, plasma confinement), galactic spiral arms (large-scale instabilities that collect material into molecular clouds), or the formation of stars or planets (pressure support against gravity, angular momentum loss).  However, whether magnetic fields have a {\em dominant} role in these settings, especially during star formation, is still unclear, and so the study of magnetic fields in such objects is a topic of active research.

Observationally, magnetic fields are difficult to measure, and no single technique can simultaneously give the full three-dimensional field strength and orientation.  However, significant partial information can be obtained with certain techniques.  Polarimetric photometry gives the magnetic field orientation projected onto the plane of the sky, but without direct information on the field strength or the line-of-sight contribution \cite[e.g.,][]{a96,smi00,atk04}.  In contrast, Zeeman spectral line measurements give the line-of-sight field strength (a one-dimensional projection), but are much less sensitive to the orthogonal components' strength or orientation \citep[see review by][]{cru12}.

Both techniques are technically challenging, since in both cases magnetic field measurements depend on obtaining accurate values for the polarised contributions to emission or absorption (e.g., the Stokes parameters $Q$, $U$, and/or $V$), which are usually quite small compared to the total intensity $I$ (although scattering in the near-infrared can give polarisations up to 100\%).  In the case of polarimetry, there is the additional complication of identifying the source of the polarisation, which (depending on wavelength and particle size) can be due to Rayleigh, Mie, or electron scattering, or dichroic absorption and/or emission from dust grains.

Thus, experiments capable of yielding high quality, spatially resolved images of either magnetic component (i.e., plane of sky or line of sight) are highly prized, not only because they are still rare, but also since some measure of the three-dimensional magnetic field structure can be recovered from such data by context.

As part of a multi-source campaign to better understand the role of magnetic fields in astrophysical disks in all settings, we report here one of the first results from the newly-commissioned mid-infrared (MIR) \cc\ instrument, a multi-mode user facility for the Gran Telescopio Canarias (GTC). 
We used \cc\ in its imaging polarimetry mode with a series of MIR filters to observe the polarisation of the 10\,$\mu$m silicate feature towards the massive star-forming region K3-50, which \citet{w07} selected as an ideal candidate for high-resolution MIR polarimetry.  Our intent was to use these observations to construct, from the dichroic properties of silicate grains in this source, one of the first examples of a spatially-resolved map of the magnetic field structure, from both emitting and absorbing layers along the line of sight and across the field of view.  We wanted to use this ``polarisation tomography''  to elucidate the nature and configuration of this source's magnetic fields, and determine whether any detected fields in the disk, outflow, or embedded objects were playing a dominant role (or not) in the evolution of this region.

K3-50 is part of the radio source W58, located in the outer Galaxy at ($l$,$b$) = (70\fdeg3,+1\fdeg6).  The distance to K3-50 has been taken as 8.7\,kpc in most recent work, as originally determined by \citet{h75}.  However, \citet{p03} pointed out that this should be rescaled to 7.4\,kpc 
with a value of $R_0$ = 8.5\,kpc for the Sun's Galactocentric distance, and proportionate rescalings would apply for other values \citep[e.g., $d$ = 7.3\,kpc with a more modern $R_0$ = 8.4\,kpc;][]{r09}.  While acknowledging this, we nonetheless adopt $d$ = 8.7\,kpc 
to be consistent with \citet{hkp97} in particular, so that we can compare our results with their model for the molecular torus (\S\ref{torus}).  A 16\% smaller distance will obviously reduce all masses and luminosities discussed herein by 29\%.

K3-50 is an archetypical luminous ($\sim$2$\times$10$^6$\,L\solar) compact HII region with ongoing massive star formation, and has been observed by many workers at multiple wavelengths.  For the purposes of this study, relevant prior results include the 2\,cm continuum and recombination line interferometric study of the larger area around K3-50 by \citet{dp94}, showing the existence of a narrow, ionised, bipolar outflow emanating from the central ultracompact HII region (UCHII) K3-50A; the 3\,mm molecular interferometry maps of \citet{hkp97}, showing an inclined toroid or disk of size $\sim$0.4\,pc with a central cavity coincident with the K3-50A UCHII peak; and the near-IR speckle interferometry of \citet{hbk04}, showing a number of bright knots inside a biconical nebulosity at the centre of the molecular toroid's cavity.  In addition, \citet{oky03} obtained MIR imaging and spectroscopy of the HII region's ionising sources in the central cluster.

Because of these geometrical features and its brightness at MIR wavelengths, K3-50 makes an ideal object for polarisation mapping of its magnetic field configuration in the 10\,$\mu$m silicate feature.  Besides the new data and analysis presented here, the only other published IR polarimetry of K3-50 IRS were presented by \citet{dc78} and \citet{smi00}, with a non-detection at 100$\mu$m by \citet{n89}.  \citet{smi00} also performed a decomposition of the absorptive and emissive components at their 5$''$ resolution, and an analysis of the orientation of these components in terms of disk/toroid and outflow axes was published by \citet{w07}.

We describe the camera and observational details in the next section, and present the results of the data reduction and analysis in \S3.  We discuss the implications of these results in \S4, and end with a brief conclusion.

\section{Observations}
\subsection{CanariCam}
\cc\ is a multi-mode instrument for the GTC, offering imaging and spectroscopy 
in the mid-infrared N (7.5-13.5 $\micron$) and Q (16-26 $\micron$) atmospheric bands, and polarimetry in the N band.  Descriptions of the overall instrument design and of the polarimetric mode have been given by \citet{cc1} and \citet{cc2}; see also http://www.gtc.iac.es/instruments/canaricam/cana- ricam.php.  \cc\ is built around a 320$\times$240 Raytheon array, forming images at a pixel scale of 0\farcs08 and giving a native field of view of 25\farcs6$\times$19\farcs2.  \cc\ is equipped with a range of broad- and narrow-band filters with central wavelengths between 7.8 and 24.5 $\micron$.  The theoretical angular resolution of the camera at 10\,$\mu$m is 0\farcs25 (1.22$\lambda/D$), but the seeing at the GTC degrades this somewhat, depending on observing conditions (see \S\ref{observations}). 

During integration in the polarimetry mode, the light beam passes though a CdSe half-wave-plate (HWP) modulator, which is rotated sequentially to four orientations (0\degr, 22\fdg5, 45\degr, and 67\fdg5), to allow for full polarisation calibration during data reduction.  Then, a CdSe Wollaston prism is used to separate the main optical beam into two orthogonally polarised ones (i.e., the $o$ ray and $e$ ray).  At the stop of each HWP orientation, a frame containing both $o$ ray and $e$ ray images is recorded simultaneously.  Depending on the integration time, several tens to hundreds of frames are obtained, and saved in a multi-dimensional FITS file at the end of each integration.

This dual-beam configuration minimises the effect of seeing and ``common-mode noise,'' such as that from atmospheric variations, and is the first ever dual-beam MIR polarimeter, offering significant gains in sensitivity over previous instruments (over and above being on a 10\,m telescope).  To prevent overlap between the $o$ ray and $e$ ray images of an extended object, a focal plane mask is used, and the effective field of view reduces to three narrow rectangular apertures, each of which is approximately 25\farcs6$\times$3\farcs0.

\begin{figure}
\centerline{
\includegraphics[angle=0,scale=0.56]{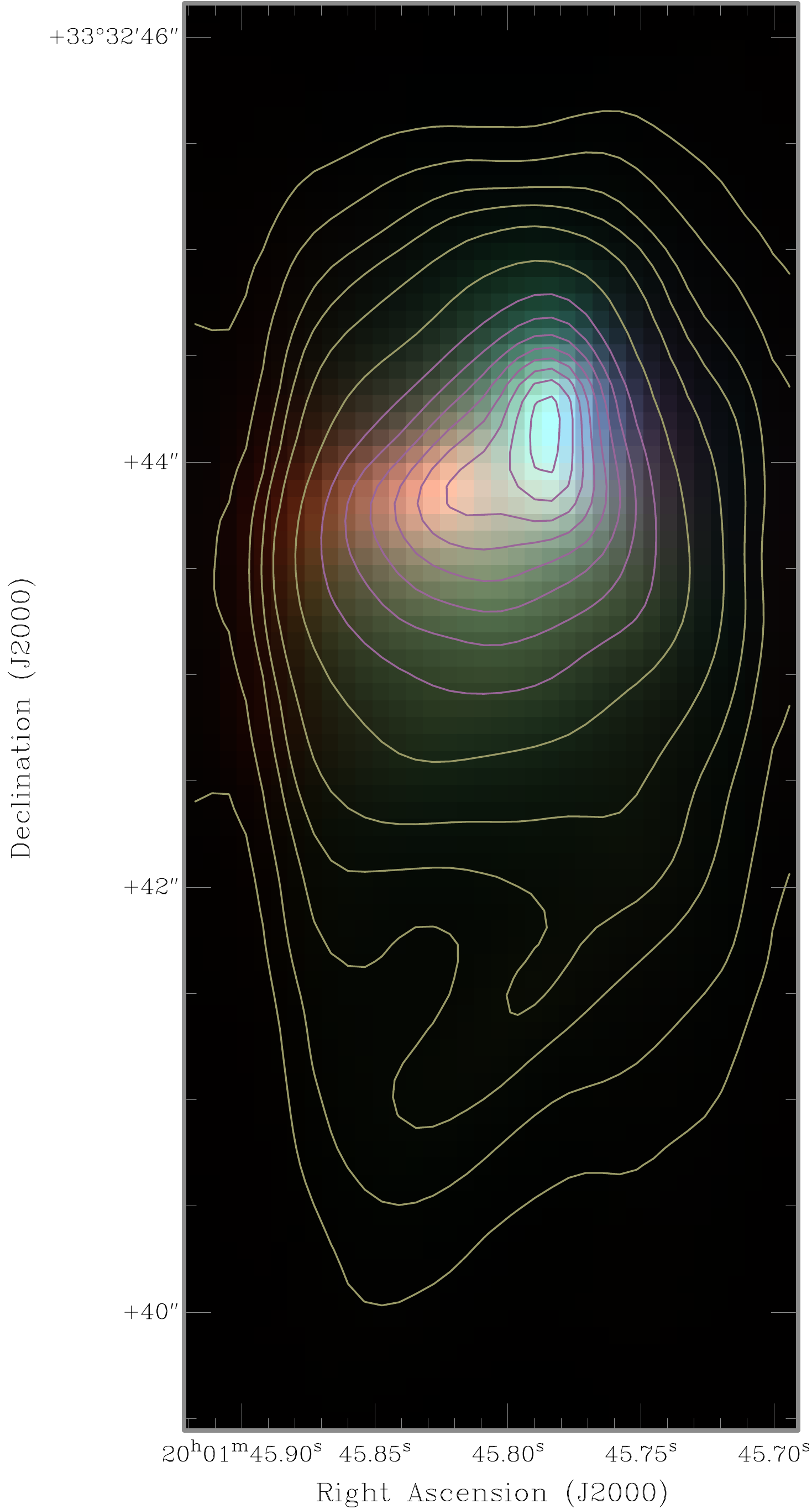}
}
\caption{Colour composite image of the 8.7\,$\mu$m  (Si2, blue), 10.3\,$\mu$m  (Si4, green), and 12.5\,$\mu$m  (Si6, red) $I$ images.  The brightness scales in each $I$ image have been adjusted to maximise the colour contrast in this image, which then represent the {\em relative} variation in flux ratios between the 3 filters.  Thus, the eastern peak has relatively higher Si6/Si4 and Si6/Si2 ratios than other parts of the image, and so appears redder; the western peak is relatively brighter in Si2 \& Si4 compared to the rest of the image, and so appears as a cyan colour; the broader area between and around these peaks has a greener hue, indicating relatively high Si4/Si2 and Si4/Si6 ratios in this area; and the far western edge of the western peak is almost pure blue, indicating relatively high Si2/Si4 and Si2/Si6 ratios compared to other areas.  These colours/ratios persist to fainter flux levels as well: red ratios to the east, cyan to the north, blue to the west, and green to the south including the ``hook'' feature.  Contours in percent of the peak of the Si4 image are shown at 0.8, 1.5, 2.6, 3.8, 6, 12\ (pale yellow), and 25 to 95 by 10 (magenta).
}
\label{colour}
\vspace{-1.5mm}
\end{figure}

\subsection{K3-50 Data Collection, Calibration, and Reduction\label{observations}}
We observed K3-50 on UT 2012 July 6 and 8 as part of the \cc\ Science Team guaranteed time for the GTC.  We positioned the \cc\ field of view over the brightest part of the near-IR speckle image of \citet{hbk04}, oriented with the detector's long axis exactly N-S. 
The three filters Si2, Si4, and Si6 were used, centred at 8.7, 10.3, and 12.5\,$\micron$ with bandwidths 1.1, 0.9, and 0.7\,$\micron$, respectively.  This combination of filters allows us to separate the dichroic absorption and emission components from the total MIR polarisation using the methodology of \citet{atk04} (see \S\ref{aitken} below).  The observations of K3-50 were interlaced with 100\,s on-source integrations on a nearby PSF calibrator HD 168723 and a polarisation/PA calibrator AFGL 2591.  The former is a Cohen standard star \citep{c99}, while the latter is chosen from the atlas of \citet{smi00}.  The effective resolution measured on the PSF star was found to be $\sim$0\farcs3, indicating a moderate influence from seeing, but the image quality is still diffraction-dominated.

For K3-50, the total on-source integration time was 864, 437, and 728 seconds (resp.) for each filter.  The resulting rms noise in the (unfiltered, unsmoothed, unbinned, raw) total intensity $I$ images is 13, 21, and 21 mJy arcsec$^{-2}$.  We used the standard chop-nod technique to correct for the thermal background from the sky and the telescope.  The chop throw was 20$''$ along the E-W direction and the chop frequency was 2 Hz.  No flat-field correction is needed for instruments like \cc, since the pixel response is very uniform across the entire array.

We present a 3-colour composite of the raw $I$ images in Figure \ref{colour}, where one can see by the rendered colours that different structures in K3-50 have different flux ratios among the 3 filters.  In this rendering, redder colours probably indicate cooler dust and/or higher extinction, while bluer colours may indicate warmer and/or less obscured features.

\begin{figure*}
\centerline{
\includegraphics[angle=0,scale=0.62]{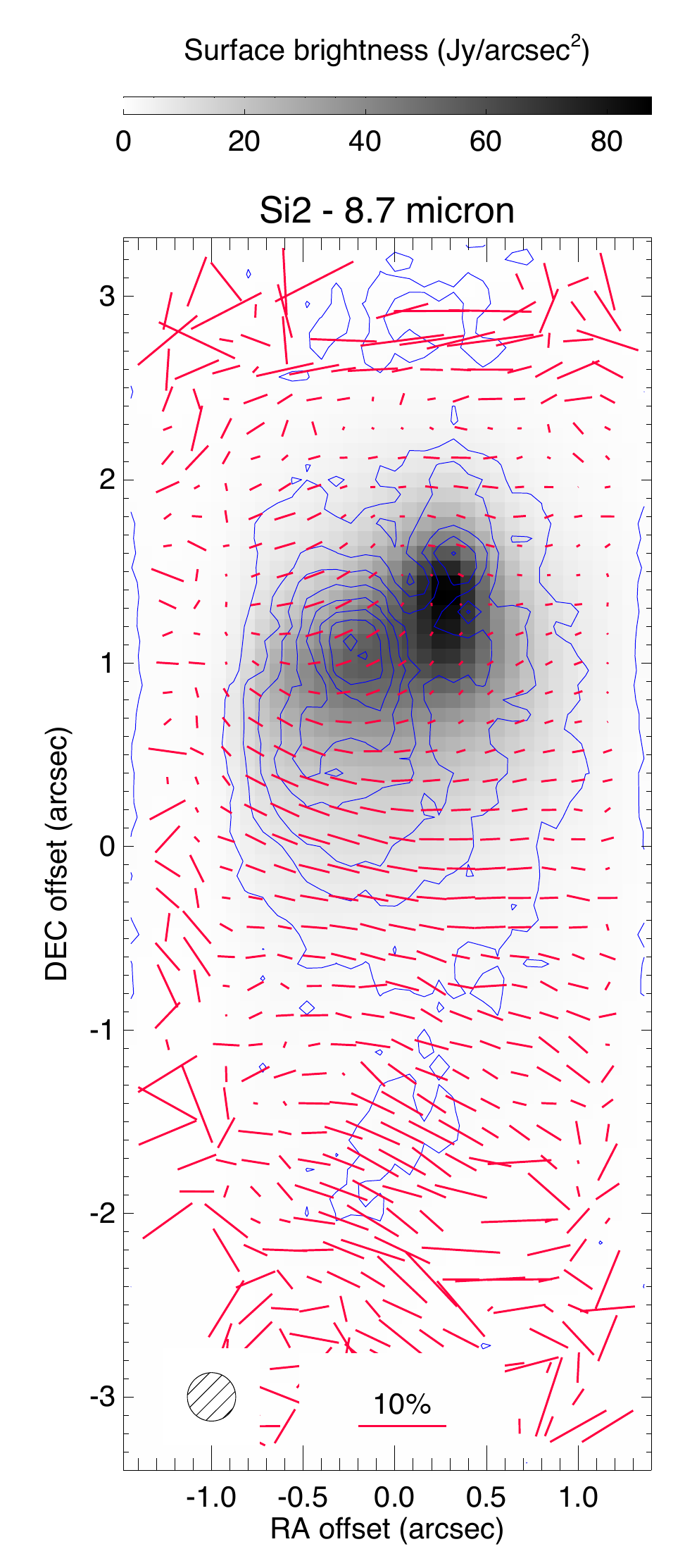} 
\includegraphics[angle=0,scale=0.62]{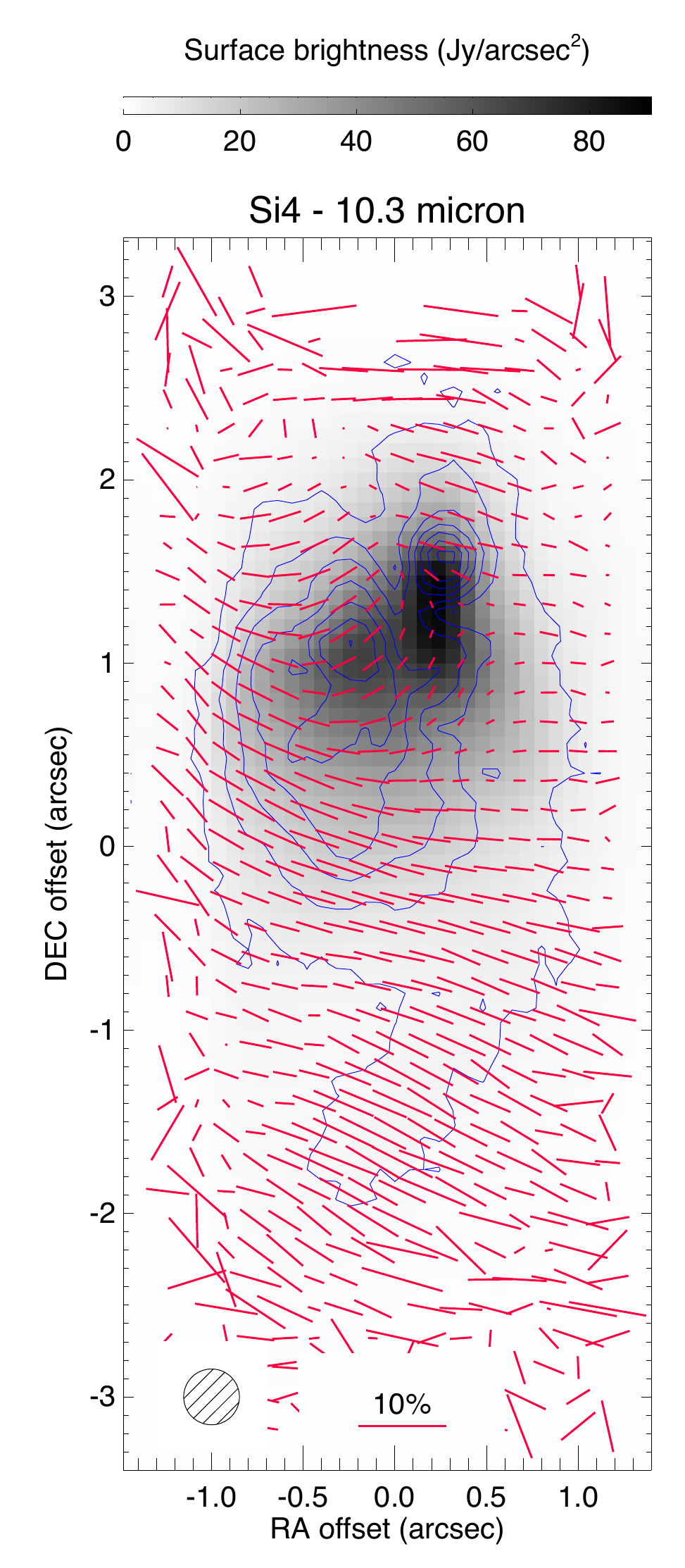}
\includegraphics[angle=0,scale=0.62]{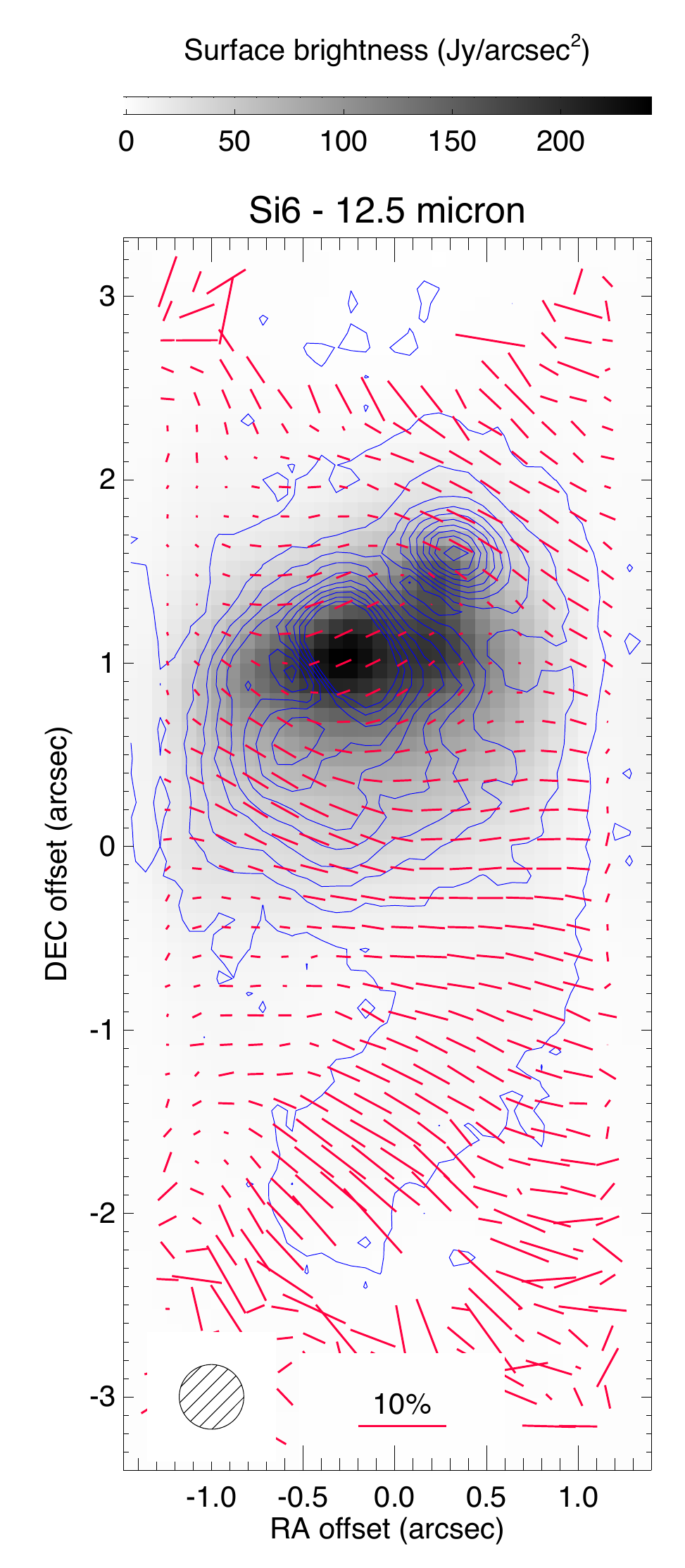}
}
\caption{\cc\ imaging products for K3-50 at (left to right) 8.7, 10.3, and 12.5 $\mu$m, derived from the corresponding $I$, $Q$, $U$ images using {\em iDealCam}.  Note that these panels are truncated in their long dimension and a S/N $\ge$ 3 clip applied to show only the high significance data.  In each panel, the greyscale image is for the total intensity $I$, in units indicated by the corresponding {\color{black}greyscale wedge} at the top; these are consistent with images presented by \citet{oky03} and will not be further discussed here.  The blue  contours are for the polarised intensity (Si2, first at 150\,mJy\,arcsec$^{-2}$, rising in steps of 75\,mJy\,arcsec$^{-2}$; Si4, 300(150)\,mJy\,arcsec$^{-2}$; Si6, 300(120)\,mJy\,arcsec$^{-2}$); and the red vectors show the polarisation orientation and percentage as indicated by the red scale bar at the {\color{black}bottom}.  The FWHM of the camera+telescope PSF (including seeing) is shown for reference as a hatched circle in each panel; these have sizes 0\farcs27, 0\farcs31, and 0\farcs35, resp.  These images have been formed with a 2$\times$2 binning from the original data, so the pixel scale here is 0\farcs16.
}
\label{polox3}
\end{figure*}

To recover the Stokes parameters from the raw data, an IDL-based custom software package {\em iDealCam} \citep{L13} was used.  The procedures are briefly summarised here.  After first-order background correction by chopping, a pair of sub-images with identical dimensions (6\farcs6$\times$2\farcs8) was extracted from each frame, corresponding to the $o$ ray image and the $e$ ray image of the source.  The sum of the two sub-images gives a measure of Stokes $I$ (total intensity), and the difference between them is a measure of Stokes $Q$ or $U$, depending on the HWP angle.  All frames in the raw FITS file are processed in the same way, and the results are averaged at the end to generate the final images of $I$, $Q$, and $U$.  The standard deviation between frames is also computed to estimate the uncertainty of each Stokes parameter.  For the purpose of data visualisation, we also calculated the (linear) polarisation percentage $P$ (defined as $\sqrt{Q^2+U^2}/I$) and the angle of 
polarisation $\theta$ (defined as $0.5\tan^{-1}[U/Q]$) for each pixel (or binned pixel to enhance the signal-to-noise ratio) and at each wavelength.  Finally, the PA calibrator was used to translate $\theta$ to the reference system with respect to the true north on the sky.  Typical uncertainties in these quantities for a 2$\times$2-binned pixel range from $\sim$0.1 to 0.5\% in $P$, and $\sim$5--15\degr\ in $\theta$, for the brightest and outlying areas of emission, respectively.  These products are computed in {\em iDealCam} by formal propagation of measurement uncertainties in $I$, $Q$, and $U$, and are shown in combined form in Figure \ref{polox3}.

Since \cc\ is installed at the Nasmyth focus of the telescope, the tertiary flat mirror introduces significant ($\sim$0.6\%) instrumental polarisation (P-inst).  This instrumental effect has been characterised during the commissioning of \cc, and its behaviour is well known, so P-inst was 
corrected during data reduction, including allowing for the different parallactic angle between the source and PA calibrator.  A detailed description of the data collection and data reduction procedures, including P-inst correction, will be found in Li et al. 2015 (in prep.).

\section{Data Analysis}
\subsection{Polarisation Components --- The Aitken Method\label{aitken}}

Aitken and co-workers have shown that the 10\,$\mu$m spectral signature of silicate particles provides a powerful diagnostic tool for probing the complex polarimetric properties of astrophysical environments where silicates are a significant component 
\citep{a96,smi00,atk04}.  This is 
evident in Figure \ref{polox3}, where one can see a much stronger variation from panel to panel in the polarisation maps (blue contours and red vectors) than in the greyscale $I$ images.  In essence, Aitken et al.\,showed that, for small polarisations, the observed polarisation as a function of wavelength can be approximated by a linear combination of the polarisation due to dichroic absorption and that due to dichroic emission.  They then assumed that the spectral profile $f$ of the grain polarisation obeys $f_{\rm em}(\lambda)\sim f_{\rm abs}(\lambda)/\tau(\lambda)$, and that the measured absorption profile $\tau(\lambda)$ for the BN object in Orion is typical of silicate grains in all sources. 
With these assumptions, the absorptive and emissive components reach their maximum values at 10.3\,$\mu$m and 11.5\,$\mu$m, respectively.

This method does not assess the contribution of any other polarisation processes, like MIR dust or electron scattering, to the data.  Instead, the working assumption that such contributions are small compared to these dichroic effects, is verified (or not) by the goodness of fit of the decomposition to the polarisation data.

\begin{figure*}
\centerline{
\hspace{-5mm}
\includegraphics[angle=0,scale=0.59]{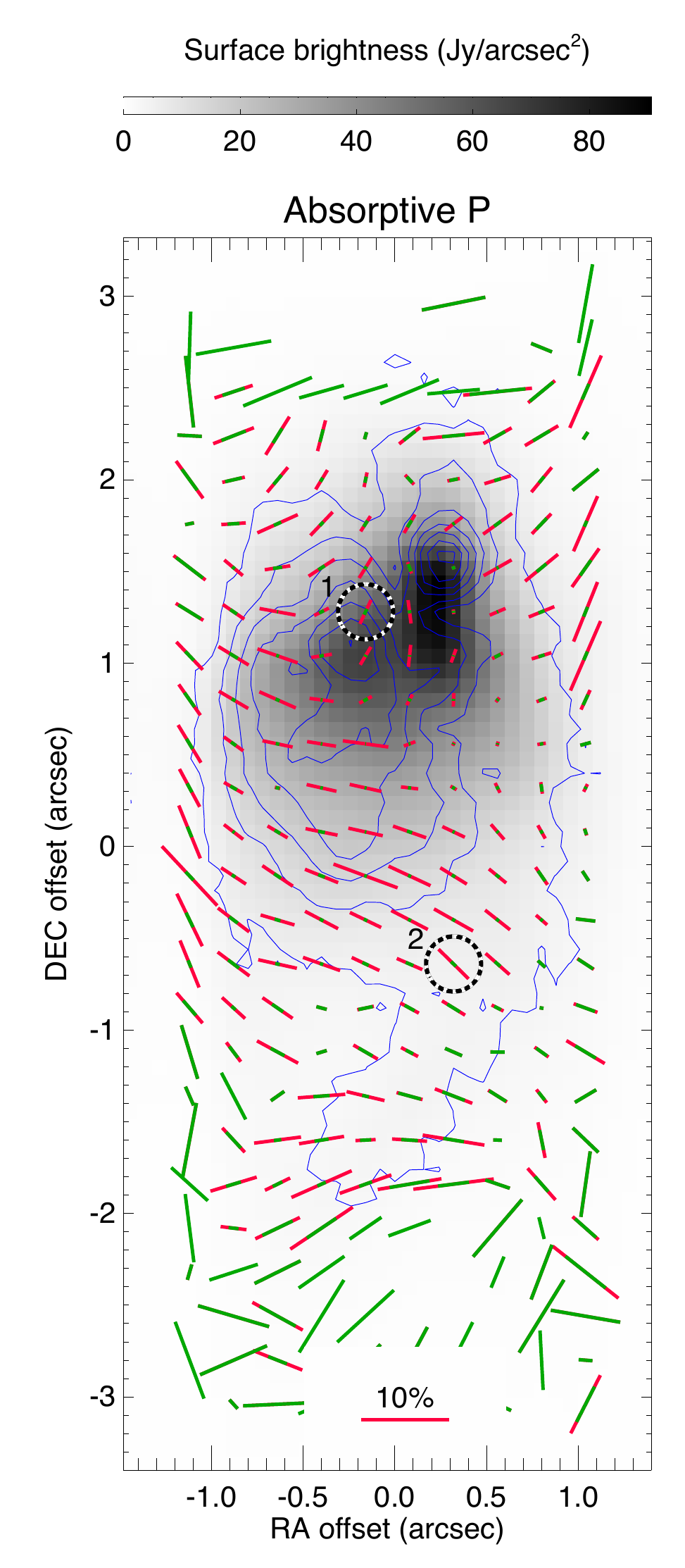}\hspace{11mm} 
\includegraphics[angle=0,scale=0.59]{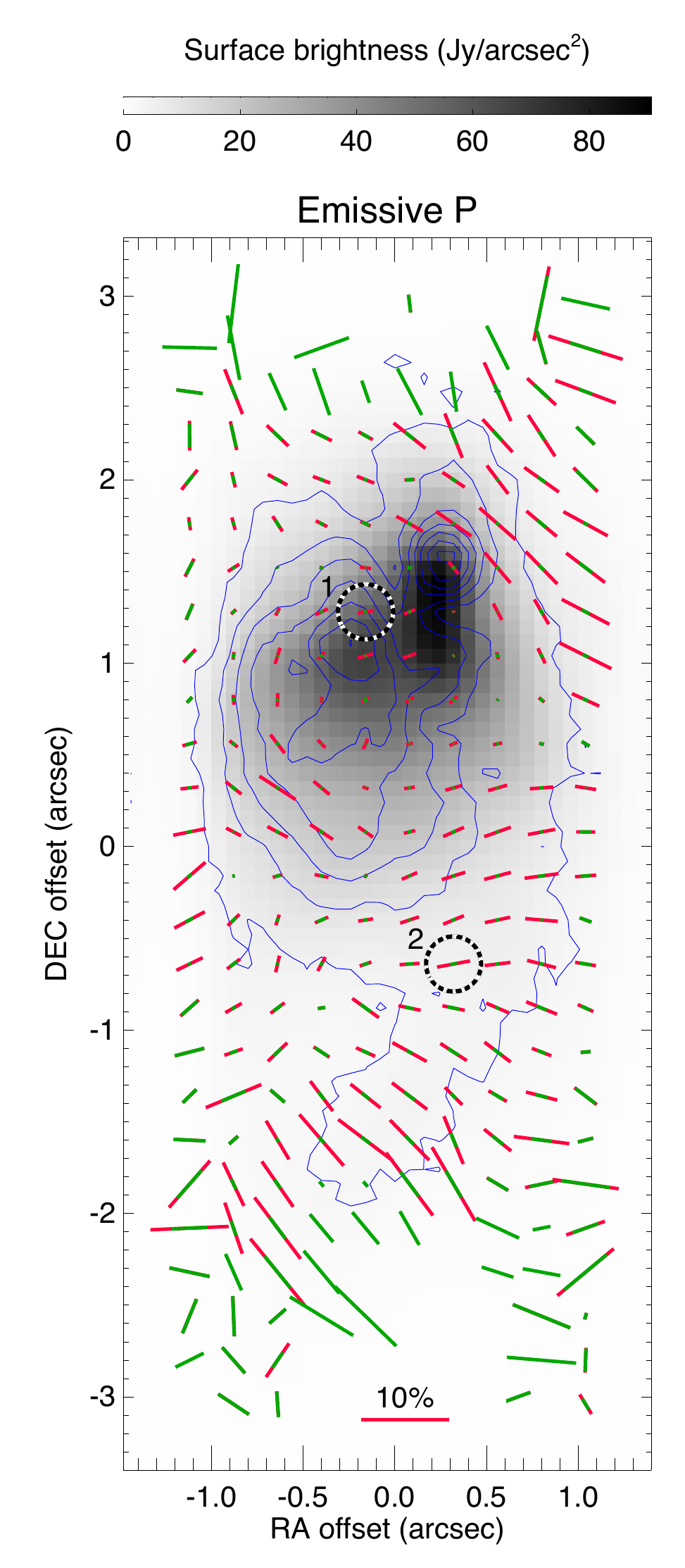}
}
\vspace{-4mm}
\caption{Aitken method decomposition of polarisation components for K3-50, now computed for each 3$\times$3 binned pixel to improve the S/N in this polarimetric decomposition. 
The absorptive component is shown in the left panel, the emissive on the right.  In each panel, the red vectors show the polarisation orientation and percentage for the respective component; the green vectors {\em within} the red vectors indicate the relative uncertainty in the polarisation percentages, but are capped to be no longer than the red vectors.  The two black-and-white dotted circles in each panel show the locations of the two sample decompositions in Fig.\,\ref{decomp}.  The absorptive polarisation reaches a maximum of 10\% at a positional offset ($\Delta\alpha$,$\Delta\delta$) near (--0\farcs1,--0\farcs2), while the emissive polarisation also peaks at 10\% near (--0\farcs2,--1\farcs8), counting only higher S/N points within the lowest blue contour.  The greyscale image, blue contours, and other details are the same in both panels, and the same as in the 10.3 $\mu$m panel in Fig.\,\ref{polox3}.  At a distance of 8.7\,kpc, the physical scale is 1$''$ = 8700\,AU = 0.042\,pc.
\label{Pcpts}}
\vspace{-4mm}
\end{figure*}

\begin{figure*}
\centerline{
\includegraphics[angle=0,scale=0.57]{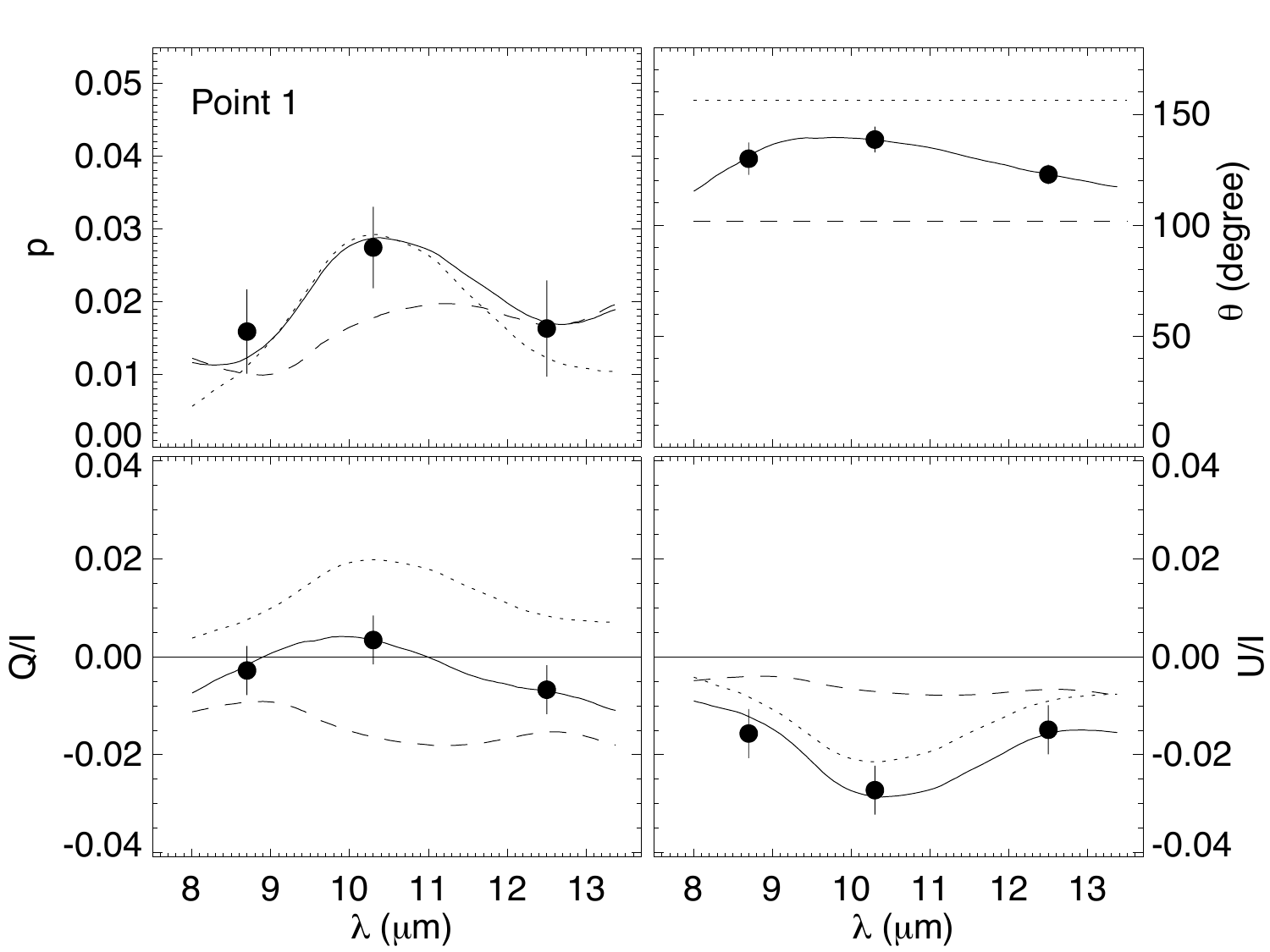}\hspace{2mm} 
\includegraphics[angle=0,scale=0.57]{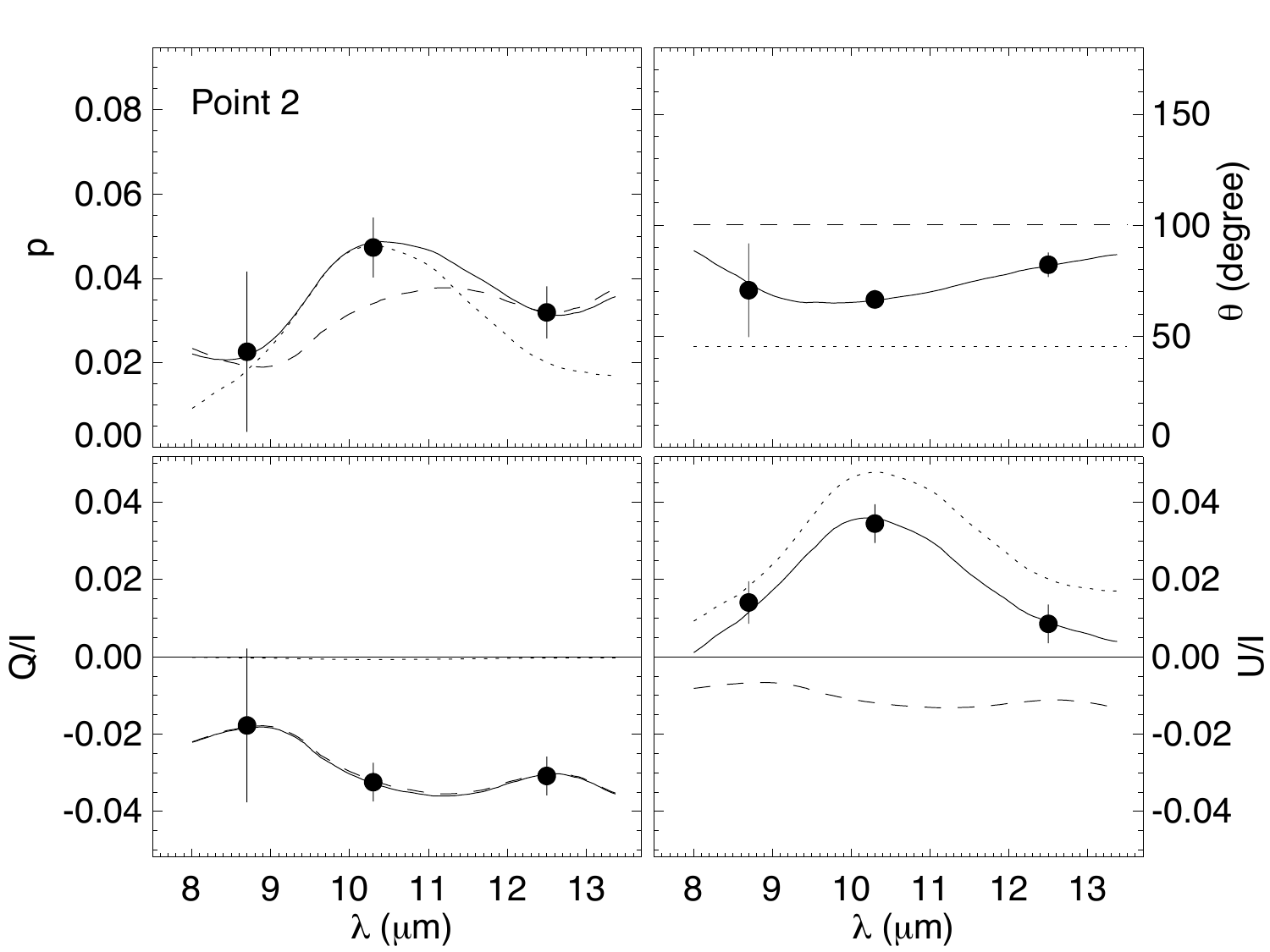}
}
\vspace{-2mm}
\caption{Sample spectral decompositions on polarimetric measurements at three wavelengths, using the Aitken method.  The two spots chosen for this illustration are at the positions indicated in Fig.\,\ref{Pcpts}, one near the central region of K3-50, and the other on the outflow cavity wall.  Similar to the presentation in \citet{atk04}, decomposition results are given in spectra of total polarization ($p$), position angle ($\theta$), Stokes $Q$/$I$ and $U$/$I$.  The solid line shows the best fit from a combination of absorptive and emissive components, which are plotted as dotted and dashed lines, respectively.  
\label{decomp}}
\vspace{-3mm}
\end{figure*}

With this approach, all that is needed to solve for the separate absorption and emission contributions is a full set of polarisation measurements for at least two wavelengths across the silicate absorption feature, as explained by \citet{atk04}.  The fact that the solutions for $P_{\rm abs}$ and $P_{\rm em}$ often successfully reproduce the observed polarisations in a variety of sources \citep[e.g.][]{smi00} underscores the reasonableness of the assumptions built into this technique (including the use of the BN profile), even when multiband polarimetry (or spectropolarimetry) is available (i.e., when the solutions are over-determined).

Using the Aitken method in {\em iDealCam}, with a functional form for $\tau(\lambda)$ given, e.g., by \citet{g75} for pure silicate grains, we derived from the data shown in Figure \ref{polox3} the corresponding solutions for the absorptive $P_{\rm abs}$ and emissive $P_{\rm em}$ components of the polarisation, after first convolving the shorter-wavelength images to the 12.5$\mu$m resolution.  These solutions are shown in Figure \ref{Pcpts} in similar format to Figure \ref{polox3}, with graphical examples of the decomposition shown in Figure \ref{decomp} for the two circled positions indicated in Figure \ref{Pcpts}.  Note that these solutions show large, coherent structures across the field of view, over a scale of many independent resolution elements.  Such features should be indicative of the inherent properties of the source, given the comparatively low measurement uncertainties.

Similarly to \citet{smi00} and \citet{atk04}, we also experimented with fitting single (emissive or absorptive) components to the \cc\ data (as opposed to fitting both components, as described above).  Using the $\chi^2$ value of the fit as a general criterion complemented by visual examination, we found that single-component fits are favoured for only a small portion of pixels ($<$10\%).  For such pixels, replacing the results from two-component fitting with single-component results causes negligible effects on the spatial structures observed, in either the emissive or absorptive polarisation maps.  Therefore, for clarity and simplicity, only the two-component fitting results are presented here.

\subsection{Magnetic Field Orientation and Polarisation Tomography\label{tomo}}
\begin{figure}
\centerline{
\hspace{-3mm}
\includegraphics[angle=0,scale=0.65]{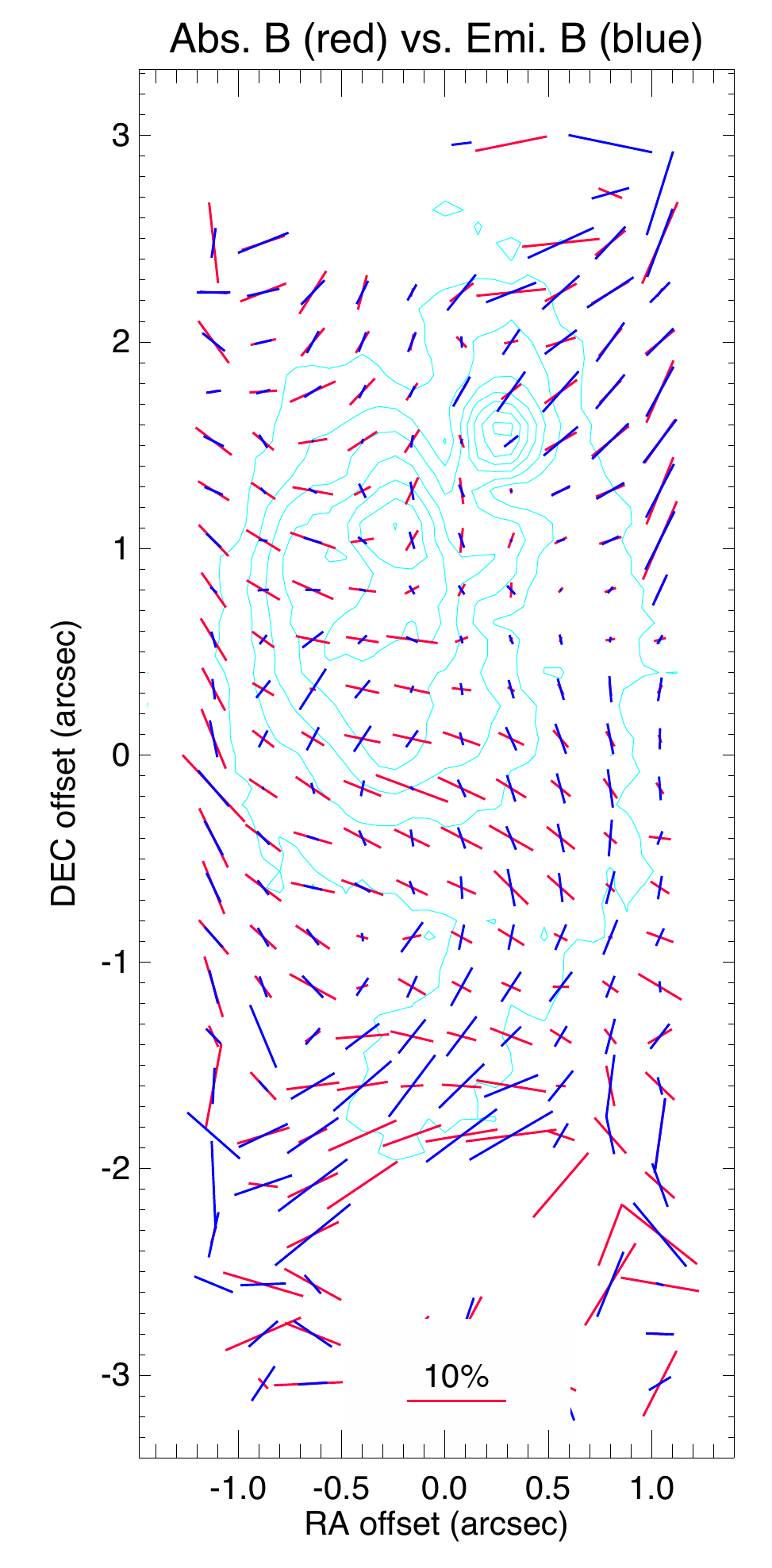} 
}
\caption{Overlay of the two components from Fig.\,\ref{Pcpts} again computed for each 3$\times$3 binned pixel.  This sampling scale (0\farcs24$\times$0\farcs24) corresponds approximately to one resolution element ($\sim$0\farcs3 including a modest seeing effect).  Here the orientation of the emission vectors (blue) is rotated by 90\degree compared to Fig.\,\ref{Pcpts}; the absorption vectors (red) are unrotated from Fig.\,\ref{Pcpts}.  The underlying cyan contours are the same as the blue 10.3 $\mu$m contours in Fig.\,\ref{Pcpts}.  The rotation of the $P_{\rm em}$ vectors means that both vector sets displayed here should represent the projected orientation of the embedding magnetic field in each of the absorbing or emitting layers, along the line of sight in each (binned) pixel.
\label{Babsem}}
\end{figure}

\begin{figure}
\centerline{
\hspace{1mm}
\includegraphics[angle=0,scale=0.6]{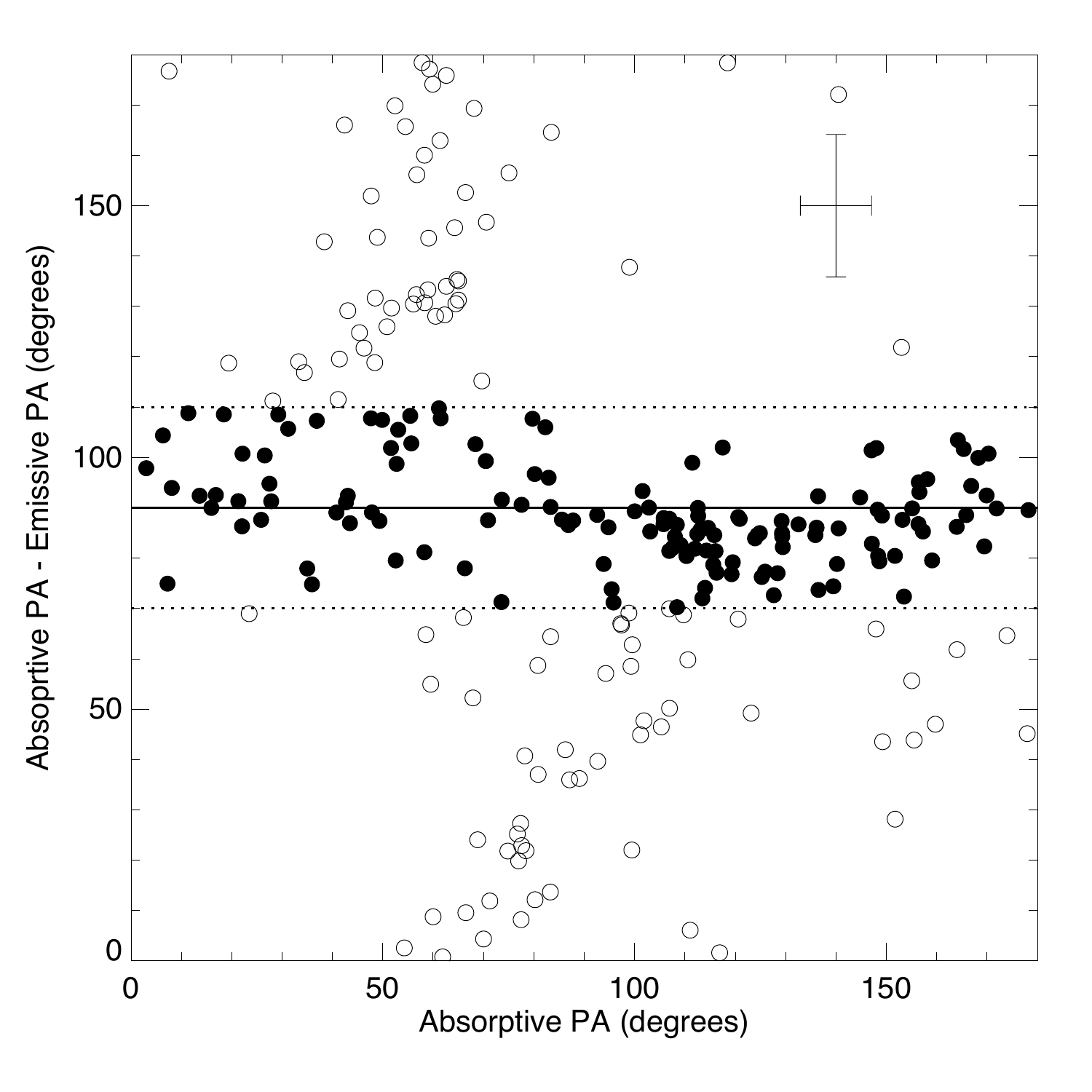} 
}
\caption{Pixel-by-pixel plot of position angle differences $\Delta\theta$ = $\theta_{\rm abs}-\theta_{\rm em}$  between the emissive and absorptive components of the silicate feature vs. position angles $\theta_{\rm abs}$ from the absorptive component.  These are the same data as shown in Fig.\,\ref{Babsem}, except that the emissive component PAs are not rotated by 90\degree\ in this plot.  The error bar shows the typical uncertainties.  For reference, a horizontal black line is drawn at $\Delta\theta = 90$\degree, where the inferred magnetic field orientations from the two components are aligned with each other.  One can clearly see that these orientations cluster into two distinct groups.  Defining $\Delta\theta_B = |\Delta\theta-90$\degree$|$, we draw horizontal dotted lines to indicate this separation, and use {\color{black}filled symbols} where $\Delta\theta_B \leq {\color{black}20}$\degree, while {\color{black}open symbols} have $\Delta\theta_B > {\color{black}20}$\degree.  In other words, {\color{black}filled symbols} have magnetic field orientations in the absorptive and emissive layers aligned to within {\color{black}20}\degree\ (hereafter the {\bf ``aligned''} points), while {\color{black}open symbols} have magnetic fields that are more poorly aligned than this (hereafter the {\bf ``unaligned''} points).
\label{absemPA}}
\end{figure}

The Aitken decomposition is significant because the absorption and emission of the polarised radiation is tied to the physical orientation of silicate grains aligned in an embedding magnetic field.  
For the absorption, the polarisation direction should be the same as the magnetic field orientation (projected onto the plane of the sky). 
Since the absorption is greater in the direction parallel to the grains' long axis, which is expected to have a net orientation perpendicular to the magnetic field, the absorptive polarisation $E$-vector will be parallel to the field direction.  
In contrast, the polarised emission that we see is preferentially oriented perpendicular to the magnetic field \citep{L07,LH14}.  Therefore, to examine the intrinsic magnetic field directions in the plane of the sky in the absorbing and emitting layers, we can overlay the two panels of Figure \ref{Pcpts}, but with the emissive vectors rotated by 90\degree.  
This is shown in Figure \ref{Babsem}, where a very interesting trend emerges on the relative orientation of the two magnetic field components. 
To quantify and further explore this trend, we plot the position angle difference between the emissive and absorptive magnetic field layers in Figure \ref{absemPA}.


From Figure \ref{absemPA} we see that there are many pixels {\color{black}(filled circles)} where the magnetic field orientation from both components is well-aligned ($\Delta\theta_B \leq {\color{black}20}$\degree).  If there are widespread areas where both the emission and absorption from silicate grains seem to trace Òthe sameÓ magnetic field, it is possible that this field is uniform across absorption and emission regions separated along the line of sight, or that these polarisation signatures are collocated or commingled, perhaps in a clumpy medium threaded by a common magnetic field.

In contrast, for areas where the polarisation indicates differently aligned magnetic fields ($\Delta\theta_B > {\color{black}20}$\degree, open circles), it is reasonable to suppose that the two magnetic field components are located in two regions separated along the line of sight.  In this case, we further suppose that the absorption-related magnetic field lies in front of the emission-related one; it is then also likely that the emission-related field, being associated with warm dust, lies closer to the main heating source for the region.

The distribution of points in Figure \ref{absemPA} show some clear ordering, and suggests that the pixels with similar magnetic field characteristics may also be spatially correlated with each other.  To examine this, we construct three subsets of the magnetic field vectors, and show each in Figure \ref{aligned}:

1. We start with that subset of the vectors from Figure \ref{Babsem} which satisfy the alignment criterion $\Delta\theta_B \leq$ {\color{black}20}\degree, indicating the location of the 
emissive and absorptive media that we postulate are threaded by the same, or nearly the same, 
magnetic field.  This first subset is shown in Figure \ref{aligned}a, and reveals two features: (i) pixels at the extreme NW and SE corners of the \cc\ frame where the field seems roughly aligned with the outflow axis from \citet{dp94}; and (ii) mid-frame pixels that seem roughly aligned with the orientation of the molecular torus from \citet{hkp97}.

There is also a narrow trough of low polarisation $P$ close to the outflow centre (shown as a long green line in Fig.\,\ref{aligned}a), which may be due to either: the magnetic field orientation being close to the line of sight near this location; an intrinsically weak magnetic field here; a field that has structure (whether ordered or chaotic) within a resolution element or along the line-of-sight; or a field that is still strong, but for some other reason the alignment is perturbed or destroyed.  The polarisation measurements alone cannot distinguish among these possibilities.

2. We next consider the magnetic field orientation for the unaligned pixels (i.e., those pixels failing the alignment criterion above; {\color{black}open circles} in Fig.\,\ref{absemPA}).  The second subset of vectors, shown in Figure \ref{aligned}b, reveals the magnetic field orientation due {\em only} to the absorptive component for these unaligned pixels.  This absorptive magnetic field is also roughly aligned with the mid-frame field in Figure \ref{aligned}a, and roughly perpendicular to the outflow axis \citep{dp94}.

Taken together, this consistency in orientation between the mid-frame magnetic fields in Figures \ref{aligned}a \& b 
strongly suggests a generally toroidally-sweeping field across the near inside surface of the NIR cavity imaged by \citet{hbk04}, and is consistent with a {\bf toroidal geometry} of similar orientation to the \hcop\ disk of \citet{hkp97}.  

3. In contrast, the vectors representing the emissive magnetic field (Fig.\,\ref{aligned}c), drawn from the same unaligned pixels as the absorptive vectors (Fig.\,\ref{aligned}b), are very different in orientation from this last subset.  Figure \ref{aligned}c shows that the magnetic field in this region is again roughly aligned {\bf along} the outflow axis, perhaps tracing the field being carried away by the ionised outflow \citep{dp94}.  In fact, the emissive magnetic field is aligned almost N-S in the inner regions, before bending to be NW-SE to the north and south.  This is very reminiscent of the morphology of the ionised outflow in Figure 2 of \citet{dp94}, which also looks N-S in the inner contours and then NW-SE in the outer contours.  Also, the emissive magnetic field is aligned almost precisely with the outflow cavity wall delineated by the X structure (or inverted-V structure) in \citet[see also Fig.\,\ref{overall}, next]{hbk04}.

These detailed features are broadly consistent with the orientation of the larger-scale (5$''$ or 0.2\,pc) components \citep{smi00}, not only in K3-50, but also in a number of other sources according to the analysis of \citet{w07}.  For example, the relatively large inclinations between the magnetic fields in the emitting and absorbing regions of large parts of K3-50, and thus probably between the outflow and torus (see \S\S\ref{torus}--\ref{outflow}), possibly explains the non-detection of polarisation at 100\,$\mu$m by Novak et al. (1989). In their relatively large beam, and with the polarisation effectively being integrated along the line of sight, the two field components would cancel each other.  This suggests an intimate connection between the magnetic fields in K3-50 and the structures revealed at other wavelengths.  We explore these relationships quantitatively next.

\section{Discussion}
\subsection{Overall Geometry}

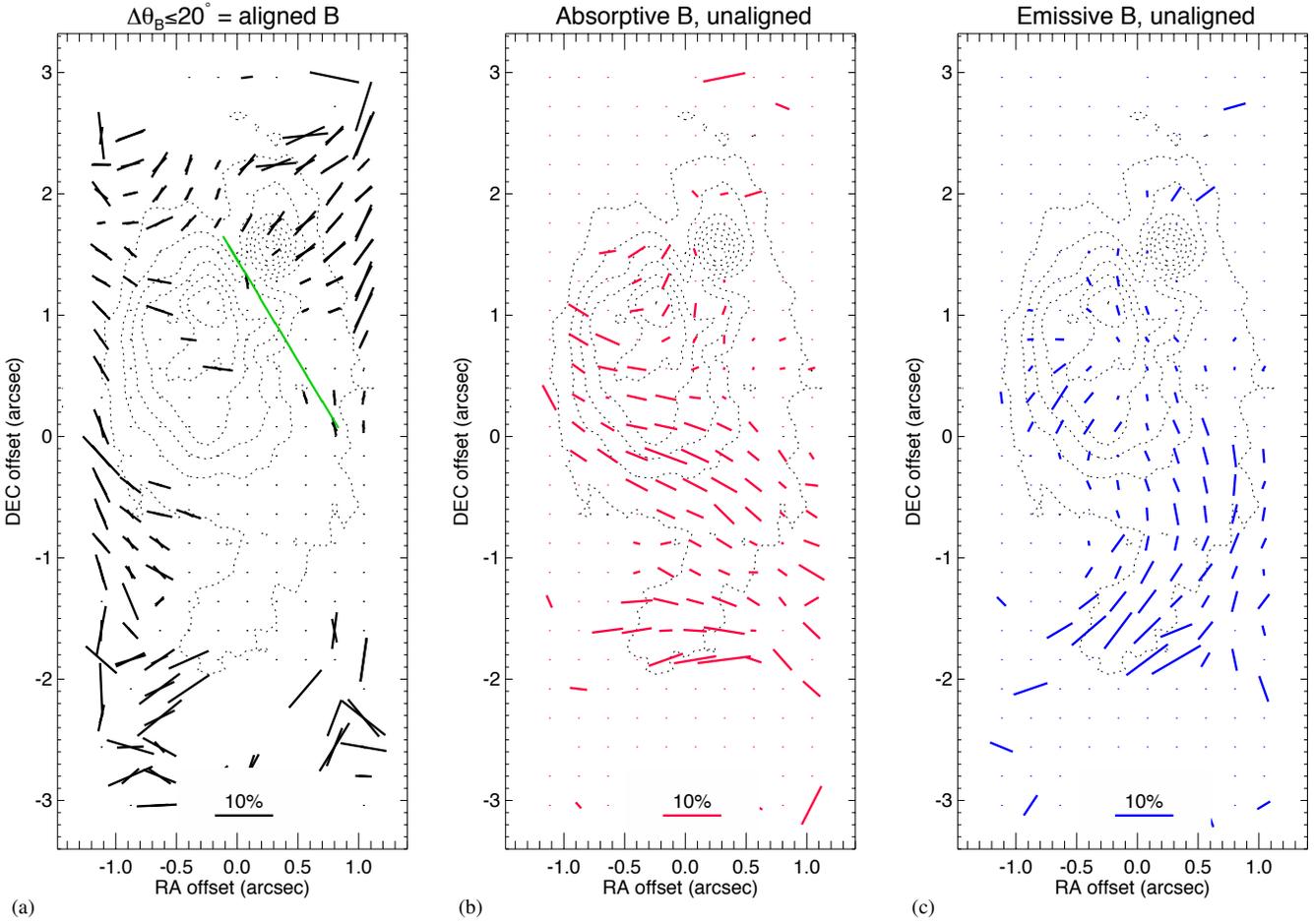
\begin{figure*}
\centerline{\hspace{3mm}
(a)\hspace{-8mm}\includegraphics[angle=0,scale=0.62]{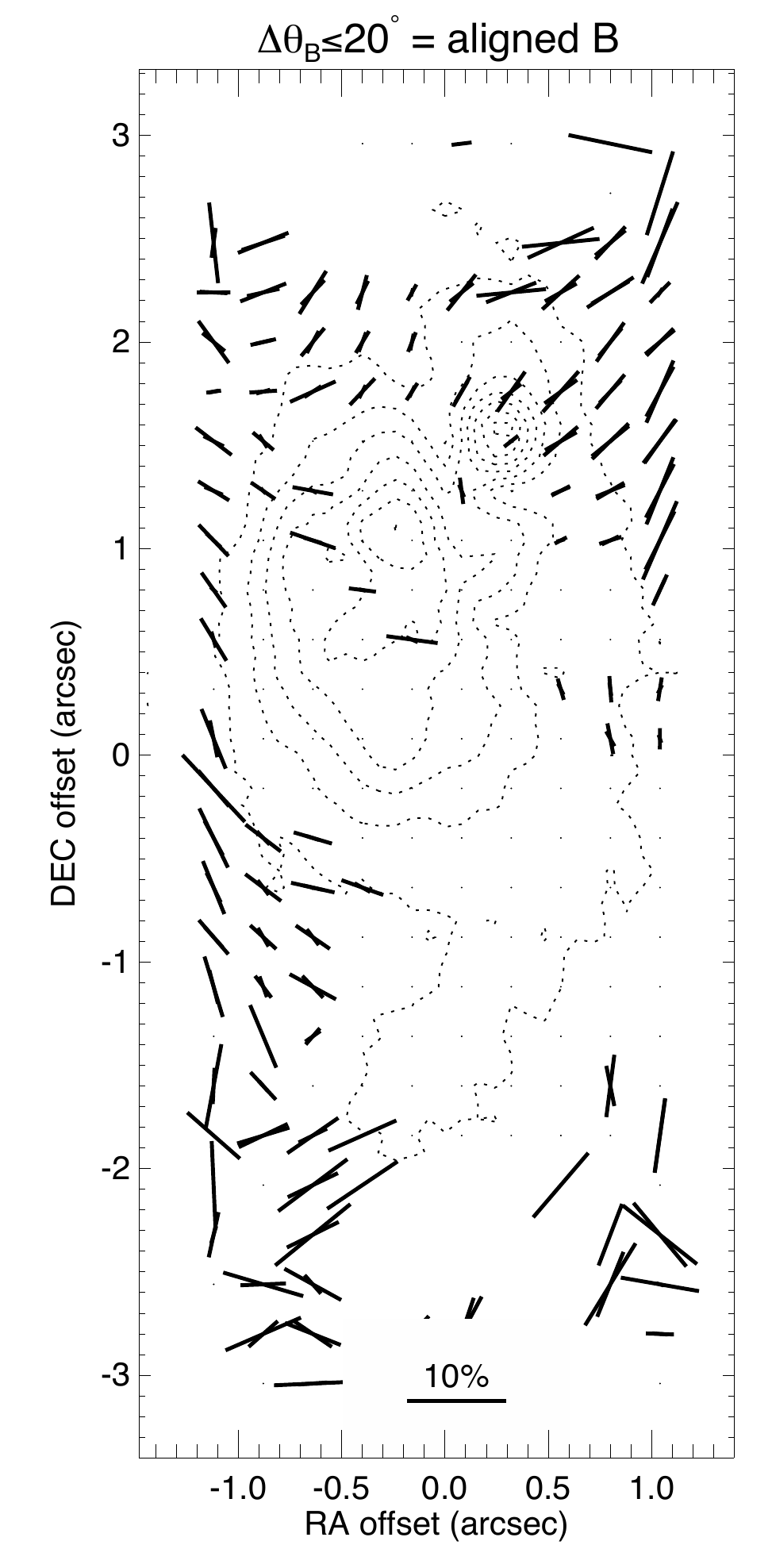}\hspace{4mm}
(b)\hspace{-8mm}\includegraphics[angle=0,scale=0.62]{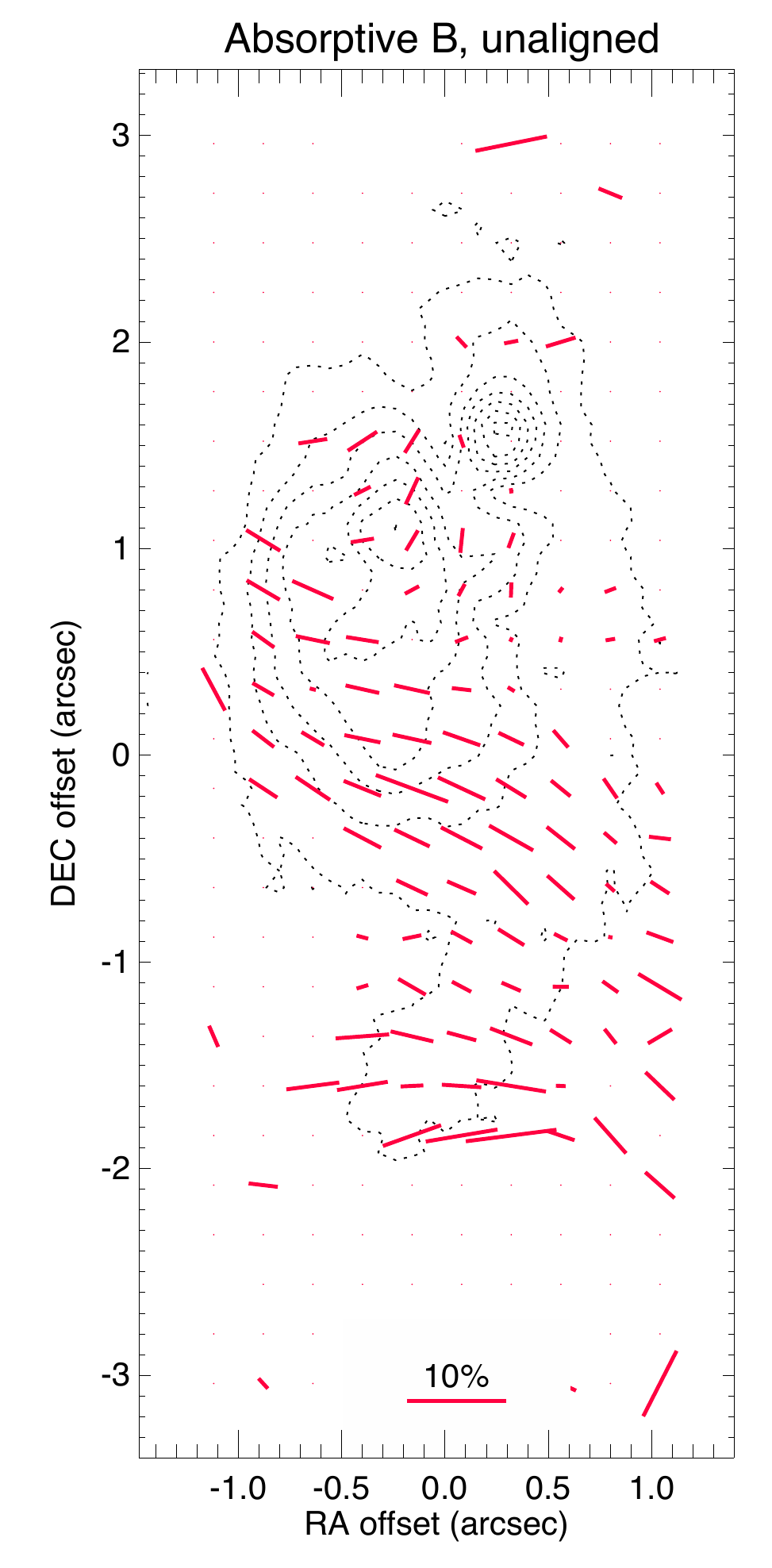}\hspace{4mm}
(c)\hspace{-8mm}\includegraphics[angle=0,scale=0.62]{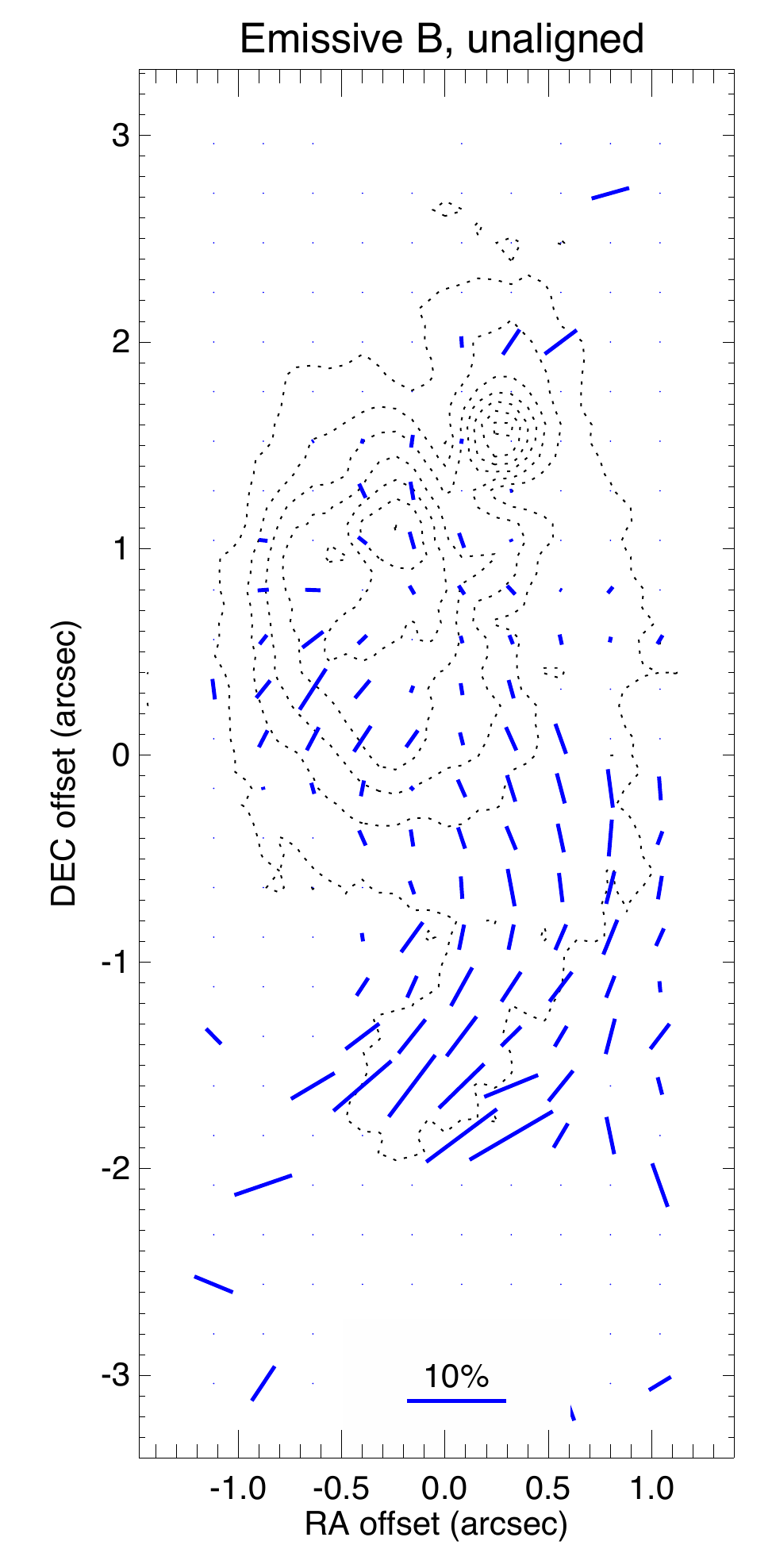}
}
\caption{Maps of magnetic field orientations.  In all 3 panels, the dotted black contours here are the same as the blue/cyan Si4 polarised intensity contours in Figs.\,\ref{polox3}, \ref{Pcpts}, and \ref{Babsem}.  (a) Magnetic field orientations from that subset of pixels in Figs.\,\ref{Babsem} and \ref{absemPA} where the field giving rise to {\bf both} the polarised absorption and emission is {\bf aligned}, i.e., $\Delta\theta_B$$\le$20\degree.  The location of a narrow trough of low polarisation $P$ close to the outflow centre is shown by a green line. 
(b) {\bf Absorptive}-only magnetic field orientations from pixels with {\bf unaligned} fields.  (c) {\bf Emissive}-only magnetic field orientations from pixels with {\bf unaligned} fields.
\label{aligned}}
\begin{picture}(1,1)
\thicklines
\definecolor{darkgreen}{rgb}{0,0.8,0}
{\color{darkgreen}\put(-125,259){\line(-3,5){44}}}
\end{picture}
\end{figure*}

Upon comparing Figures \ref{colour}--\ref{aligned}, some general trends are immediately obvious.  One is that the magnetic field structures are embedded in the diffuse emission.  The emission peaks do not seem to play an important role in distorting the field, although the western $I$-peak (cyan in Fig.\,\ref{colour}) seems to have relatively low polarisation, and so may be partially responsible (because of beam dilution) for the trough of low polarisation mentioned in \S\ref{tomo}.  The eastern compact $I$-component (red in Fig.\,\ref{colour}) seems to have a fairly constant angle of polarisation in all 3 filters (Fig.\,\ref{polox3}), and is also brightest in $I$ in the 12.5$\mu$m filter.  This suggests that it is a cold object(s) with significant silicate absorption.

The other striking thing is that the vectors in Figure \ref{aligned}a, where the emissive and absorptive components are coaligned, are confined to the outer regions of the map, and generally avoid the bright structures.  This presumably means that the dust in these regions is relatively cold and again the polarisation is probably being traced in the diffuse material.  This suggests that it is necessary to measure polarisation in the faint diffuse structures rather than the bright peaks to elucidate the dominant field morphology.

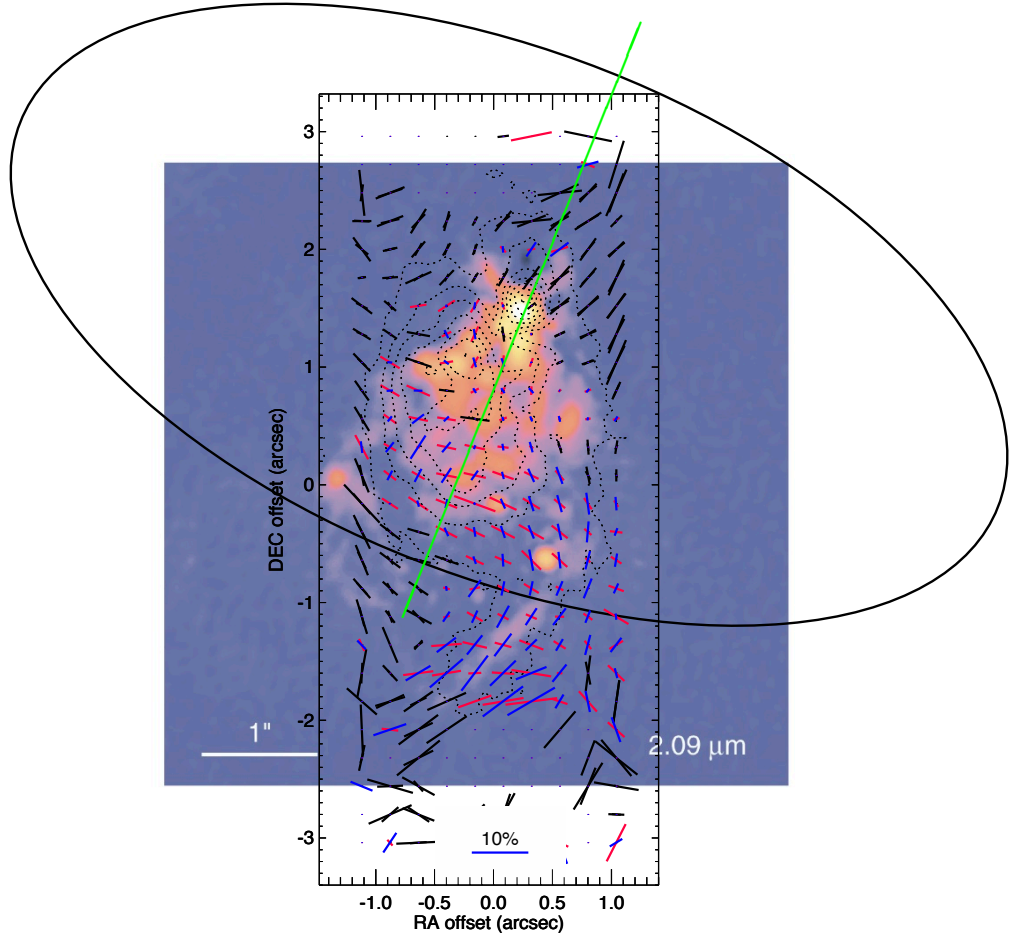
\begin{figure*}
\vspace{20mm}
\centerline{\includegraphics[angle=0,scale=0.69]{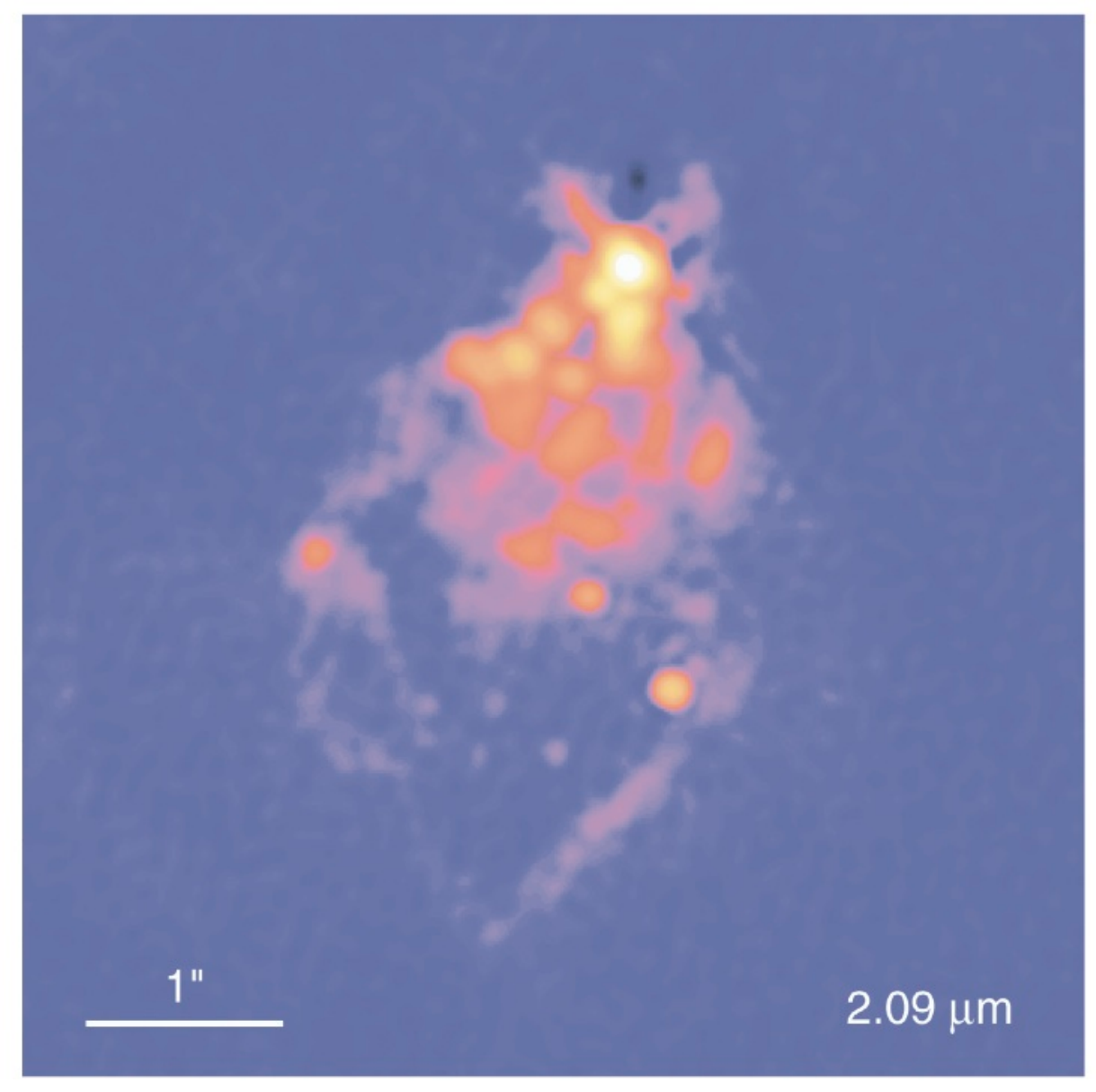}}
\vspace{-106mm}\hspace{10mm}\includegraphics[angle=0,scale=0.45]{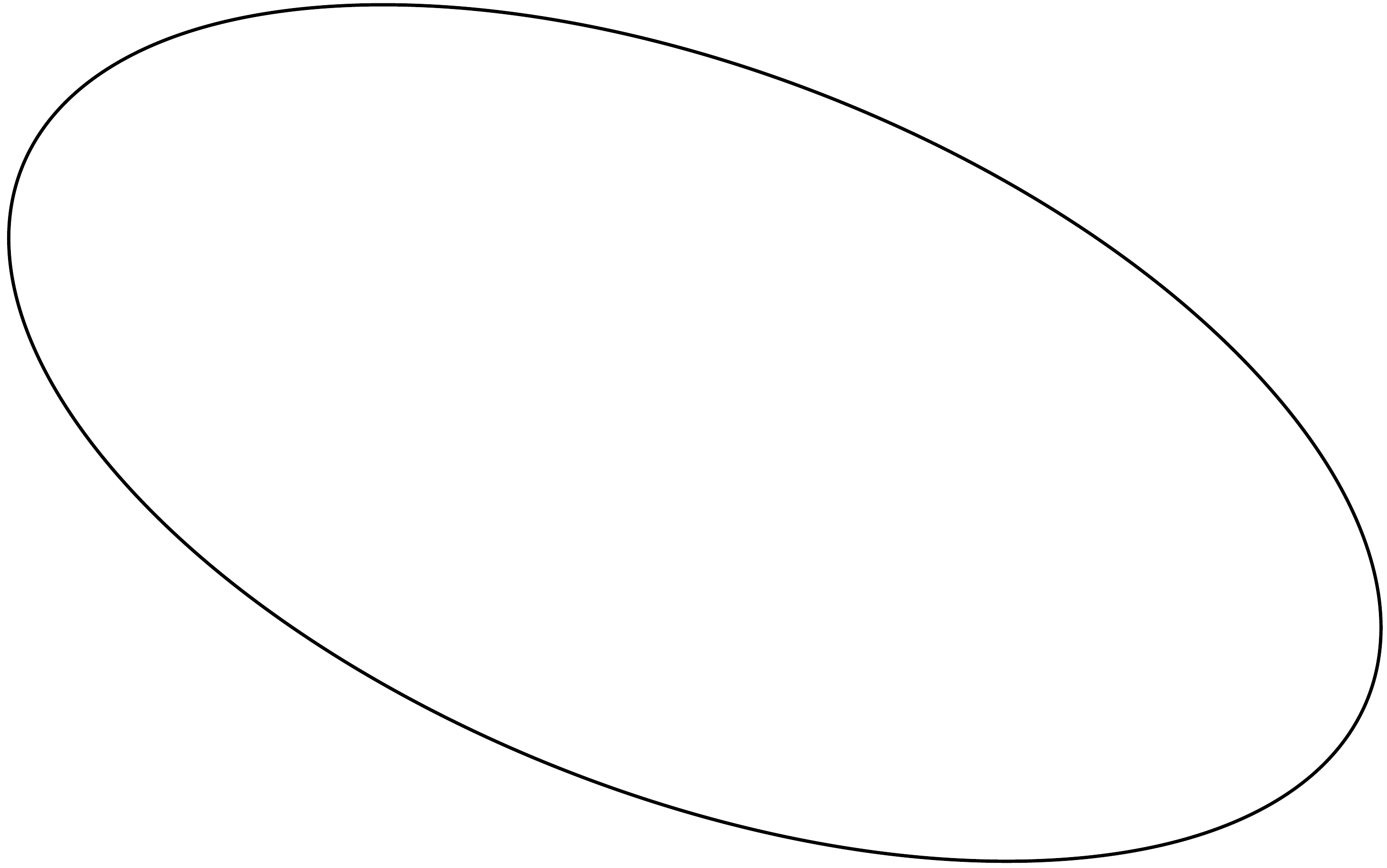}
\vspace{-77mm}

\centerline{
\hspace{-3.5mm}\includegraphics[angle=0,scale=0.592]{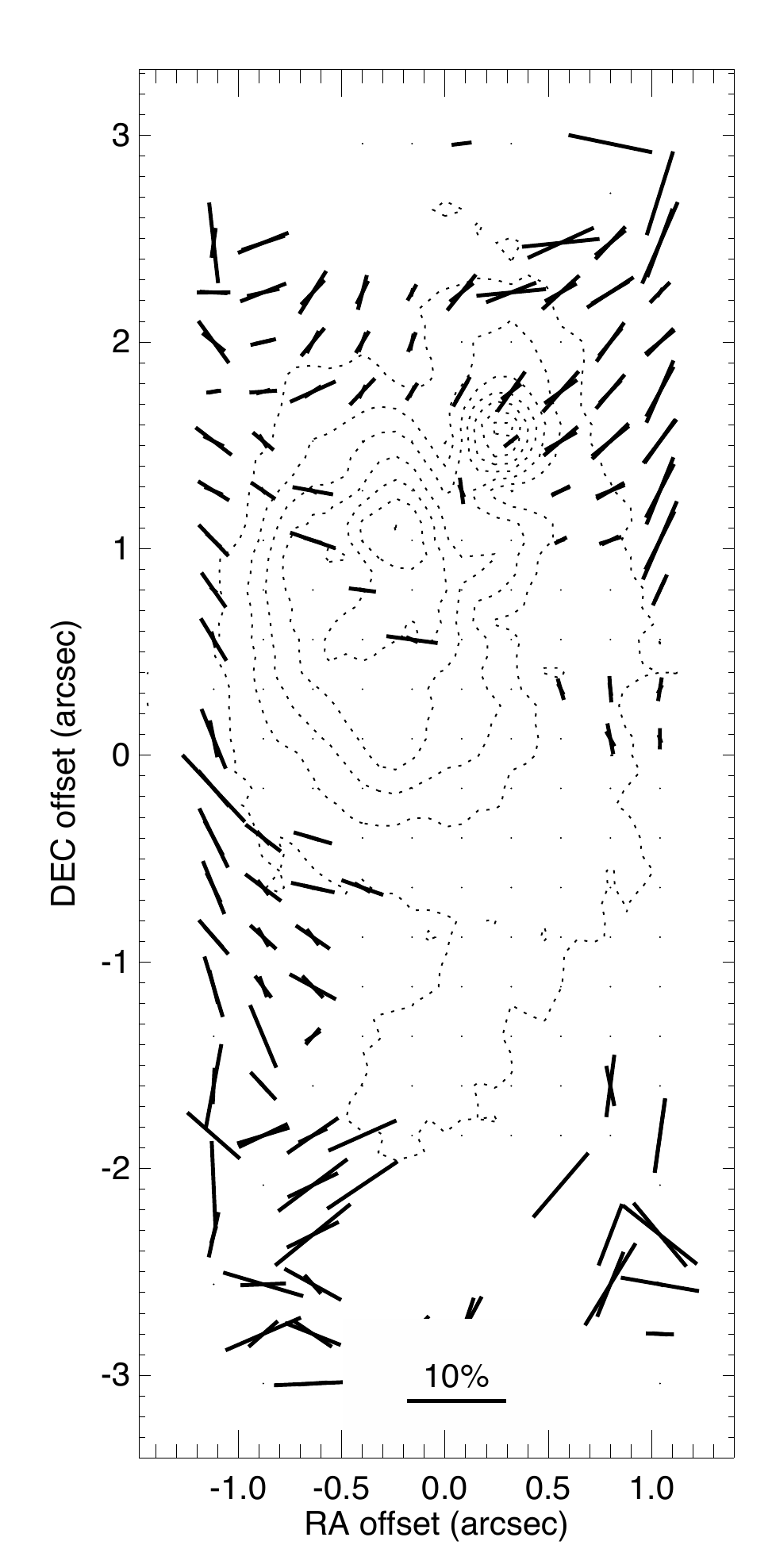}
\hspace{-59.2mm}\includegraphics[angle=0,scale=0.592]{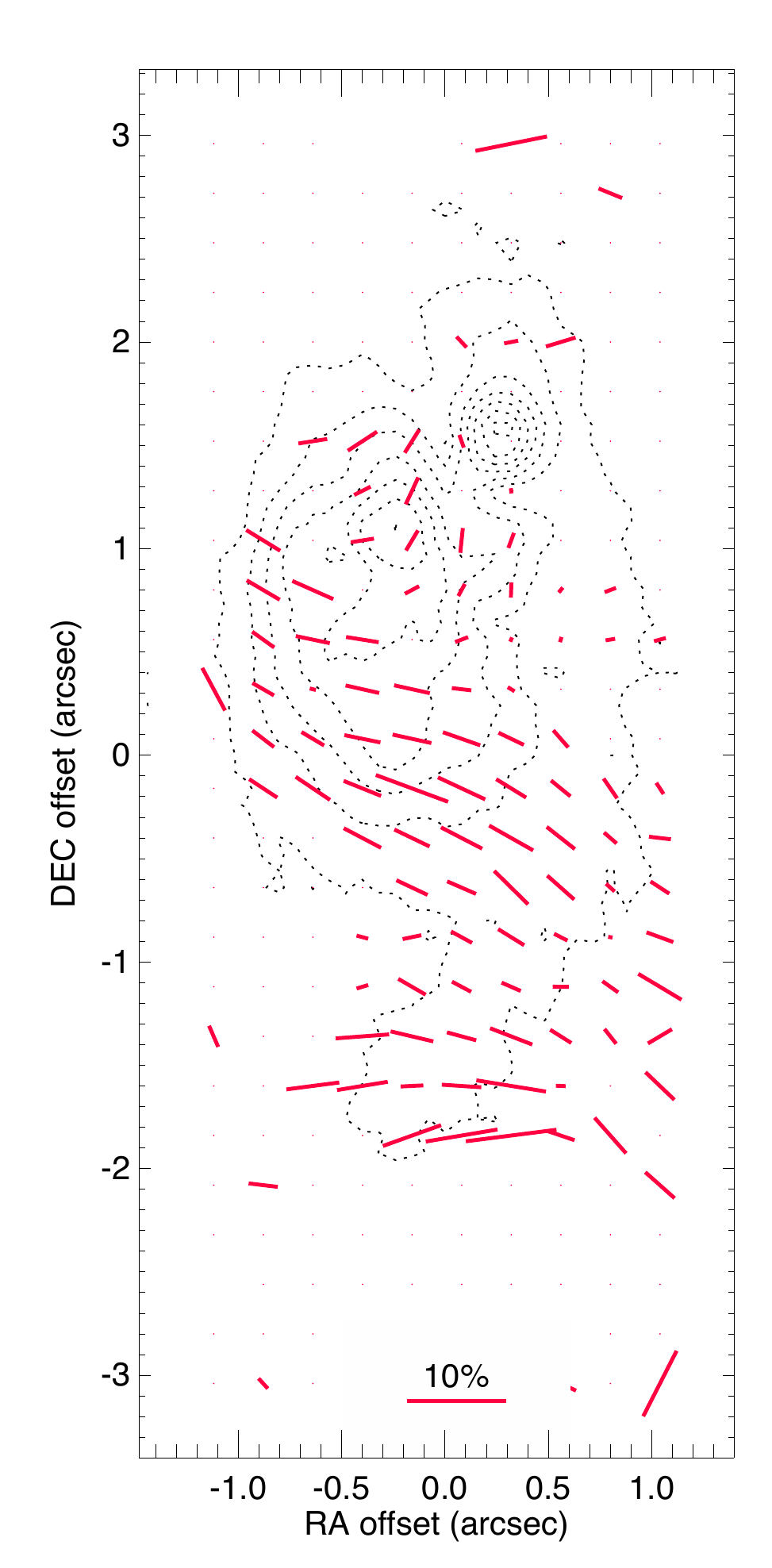}
\hspace{-59.2mm}\includegraphics[angle=0,scale=0.592]{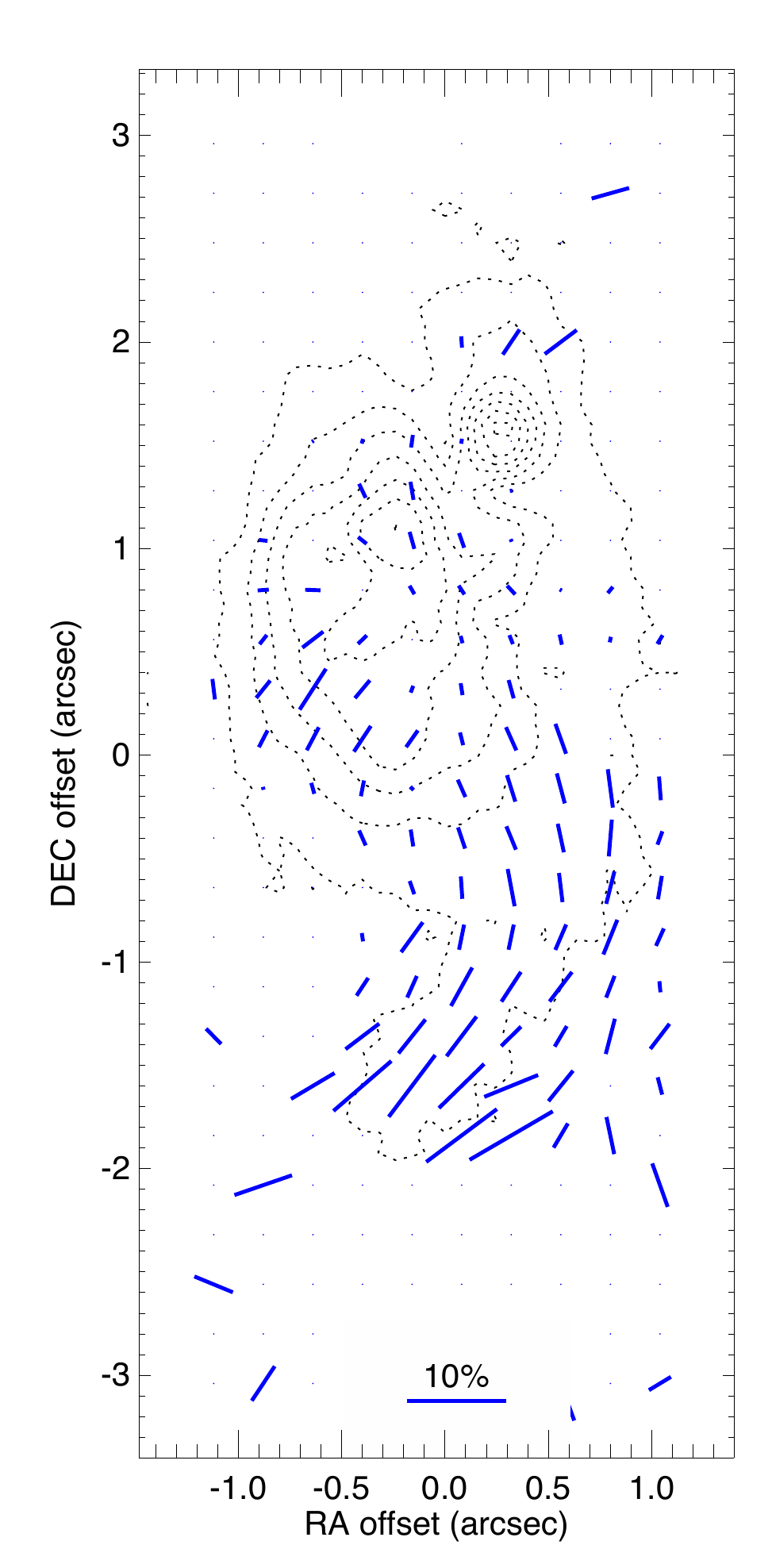}
}
\vspace{-1mm}
\caption{Main features at the centre of K3-50.  The underlying 2\,$\micron$ image is from \citet{hbk04}, and shows the possible central $\sim$0.1\,pc wide cluster of IR sources inside the southern portion of a biconical cavity (suggested by the X-shaped boundary around the IR peak) of size $\sim$0.2\,pc.  The green line indicates the approximate direction of the outflow axis, which we take as N158\fdg5E \citep{dp94,hkp97}.  The black ellipse shows the approximate size (0.44\,pc) and orientation (minor axis at --21\fdg5 from N) of the \hcop\ toroid, also from \citet{hkp97}.  The overlays of coloured polarisation vectors are the same as in Fig.\,\ref{aligned}.  At a distance of 8.7\,kpc, the scale is 20 pixels = 1\farcs6 = 14,000\,AU = 0.067\,pc.
} 
\label{overall}
\begin{picture}(1,1)
\thicklines
{\color{green}
\put(19,307){\line(2,5){45}}
\put(19,307){\line(-2,-5){45}}
}
\end{picture}
\end{figure*}

\begin{figure*}
\centerline{\hspace{-2.5mm}\includegraphics[angle=0,scale=0.45]{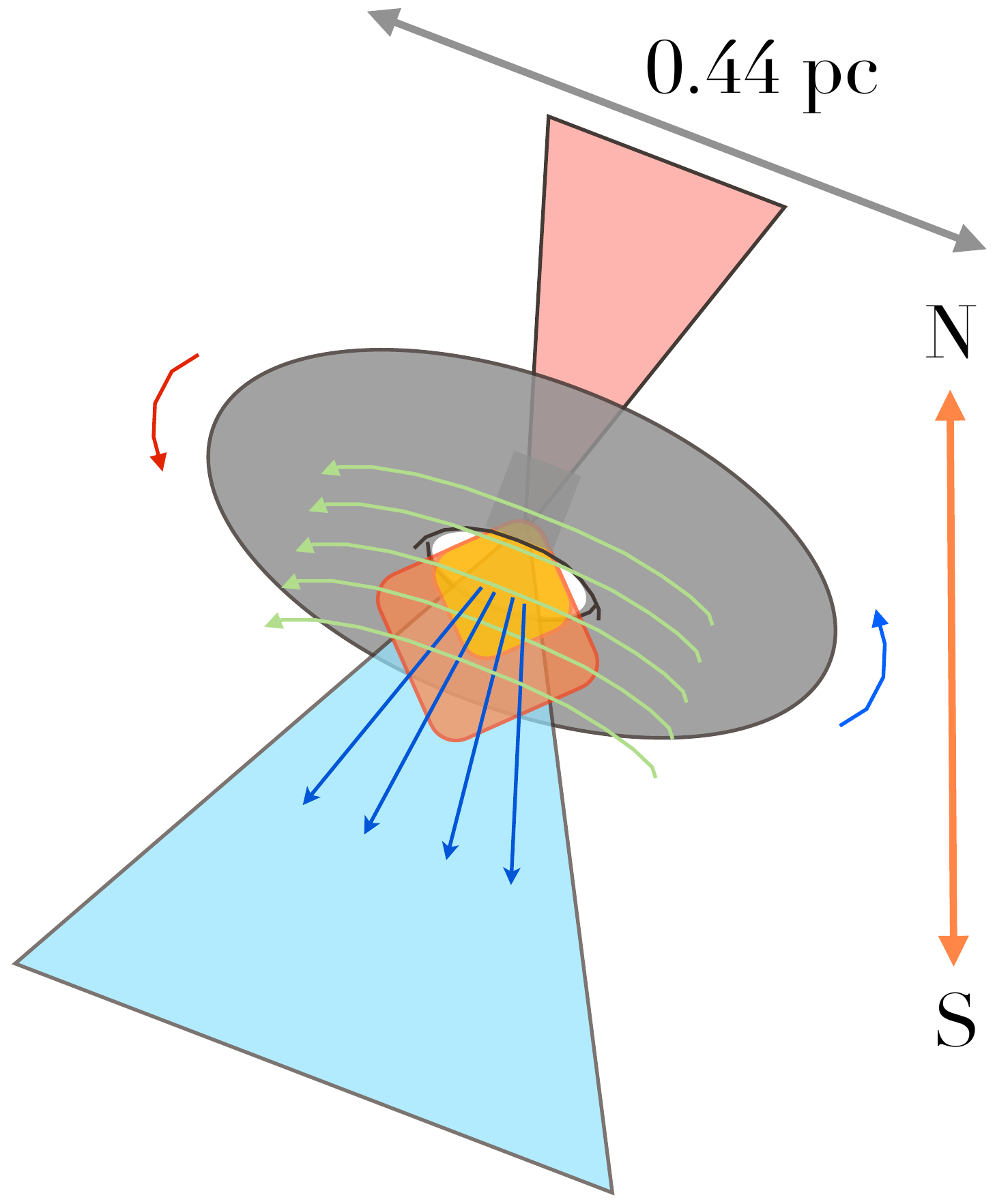}}
\begin{picture}(1,1)
\thicklines
{\color{red}
	\put(-135,235){\line(3,-2){107}}
	\put(-135,6){\line(5,3){107}}}
{\color{blue}
	\put(21,163){\line(5,3){120}}
	\put(21,71){\line(2,-1){120}}}
\put(-27,71){\framebox(45,90)}
\end{picture}\vspace{-86.5mm}

\hspace{-138mm}\includegraphics[angle=0,scale=0.45]{Baligned-pm20.pdf}
\hspace{-45.1mm}\includegraphics[angle=0,scale=0.45]{Babs-nonaligned-pm20.pdf}

\vspace{-91mm}
\hspace{132.6mm}\includegraphics[angle=0,scale=0.45]{Bem-nonaligned-pm20.pdf}
\caption{Schematic diagram of essential features of K3-50, shown as idealised geometries with approximate sizes, shapes, and orientations.  In grey we show the \hcop\ toroid from \citet{hkp97}, with the major axis size as indicated and the toroid's rotation from the \hcop\ data also shown by the red and blue arrows at the toroid's ansae.  The inner part of the IR cluster of sources from \citet{hbk04} is shown in gold, while the fainter IR cavity around this appears in orange.  Using our \cc\ data, we discern from the polarised silicate absorption a toroidally-configured foreground magnetic field (green arrows) that seems to share the toroid's geometry and perhaps also its rotation.  In contrast, the magnetic field (blue arrows) giving rise to the polarised silicate emission seems to be aligned with the blueshifted motions of the near side of the bipolar ionised outflow (light blue wedge) of \citet{dp94}.  The redshifted, far side of this outflow is partially obscured by strong IR extinction (light red wedge), presumably due to the material in the front side of the \hcop\ toroid.  The axis of symmetry of this geometry on the sky is rotated --21\fdg5 from N, as shown to the right of the diagram.  On the far left of the figure, and keyed to the diagram, we also show the vectors of the observed magnetic field associated with silicate absorption (i.e., a combination of Figs.\,\ref{aligned}a and b); on the far right we show the magnetic field associated with silicate emission (Fig.\,\ref{aligned}c).
}
\label{toymodel}
\end{figure*}

Building on the previous work of \citet{dp94}, \citet{hkp97}, \citet{hbk04}, in Figure \ref{overall} we schematically show, as a set of overlays on the \citet{hbk04} NIR image, the alignment of our polarisation maps with several of the salient features of K3-50.  Based on these comparisons, we have constructed a toy model for the arrangement of features in this source, and especially showing the relationship of the derived magnetic field configurations to the observed phenomena; this is shown in Figure \ref{toymodel}.  We note especially the alignment of the two main magnetic structures with other features thought to be common in star forming regions (namely, disks and outflows).  The toroidal magnetic field associated with polarised silicate absorption is aligned with a structure that is likely to be dominated by a large scale rotation: therefore, it suggests that the magnetic field here is also rotation-dominated (by this we mean that the rotation of the gas in the torus probably drags the field mostly along with it, and stretches the field lines enough to impart some mean field properties).  Similarly, the axial polarised-emission-related magnetic field is associated with a strong outflow of material from the central source(s) of this region, suggesting that its geometry too is determined by the outflow.  However, the opposite may also be true, i.e., that the magnetic field is actually dominant in both domains, and controlling the motions seen by \citet{dp94} and \citet{hkp97}.  Clearly, we need a way of discriminating between these scenarios: whether the magnetic fields in K3-50 are the driving agent for the structures and motions we see, or whether the fields are merely ``passengers'' as other processes control events.

\subsection{Physics of Observed Structures}

In order to determine the relative importance of the magnetic field in K3-50, we first need to ascertain the role played by other physical effects, such as gravity, rotation, thermal pressure, and turbulence.  \citet{dp94} and \citet{hkp97} provide excellent starting points for estimates of cloud mass, gravity, rotation, and other bulk parameters.

\subsubsection{The Molecular Torus}\label{torus}

We consider first the molecular torus of \citet{hkp97}.  Their self-consistent model for the torus' mass and rotation (which is apparently close to Keplerian) based on their \hcop\ data gave a best fit relation of 
\begin{equation} 
	V_{\rm rot}(R) = 2.0\kms~(R/0.10\,{\rm pc})^{0.5}  ,
\end{equation}
from which one can infer a gravitational mass contained within radius $R$ of 
\begin{equation} 
	M_{\rm rot}(<R) = V^2R/G = 93\,{\rm M}\solar~(R/0.10\,{\rm pc})^{2}  .
\end{equation}
For example, this corresponds to a mass of $\sim$450\,M\solar\ within the nominal 0.22\,pc radius of the inner torus, and a total cloud mass $\sim$5200\,M\solar\ within 0.75\,pc.\footnote{This mass is $\sim$double the value quoted by \citet{hkp97} for their integration of the same model.  The different mass estimates may be reconciled if their assumed \hcop\ abundance is made half as large.}
Because their mass estimates match reasonably well the mass profile derived from the \hcop\ integrated intensity, \citet{hkp97} concluded that the molecular torus' motions are dominated by rotation.  From the degree of mismatch between \citet{hkp97}'s model and pure rotation (mostly in the inner torus, at radii 0.1--0.22\,pc), we estimate that other effects (i.e., turbulence, magnetic fields, thermal pressure) could contribute as much as $\sim$20\% to the observed \hcop\ kinematics, if one assumes that the \citet{hkp97} model is accurate.  We explore these contributions next.

\citet{hkp97} estimated the gas temperature in the torus to be 45\,K.  We compare the contribution of thermal energy in the molecular gas to the above estimates, via the Virial Theorem:
\begin{equation} 
	M_{\rm th} = \frac{5kTR}{m_{\rm H_{2}}G} = 22\,{\rm M}\solar~\left(\frac{T}{45\,{\rm K}}\right)~\left(\frac{R}{0.10\,{\rm pc}}\right)  .
\end{equation}
Again, at a radius of 0.22\,pc, this gas temperature would provide a thermal energy content sufficient to balance the gravitational self-energy of a $\sim$47\,M\solar\ gas cloud.  Since this is much less ($\sim$25\% at 0.1\,pc, $\sim$10\% at 0.22\,pc, $\sim$3\% at 0.75\,pc) than the mass inferred from rotational-gravity balance at this radius, it indicates that thermal energy in the gas contributes a relatively small fraction to the energy content (or dynamics) of the cloud in the 
torus, and that this fraction drops off to even smaller contributions at larger radii, assuming the temperature is constant.  If the gas temperature actually drops with radius, as seems likely, then the thermal contribution to the energy content is even less.

Similarly, the Virial term giving the mass in the torus that can be supported against gravity by magnetic pressure is
\begin{equation} 
	M_{\rm mag} = \sqrt{\frac{10\pi}{9\mu_0 G}}~BR^2  .
\end{equation}
At this point, we have no estimate of the magnetic field strength\footnote{We use SI units throughout this paper, including tesla instead of gauss for the magnetic field strength $B$.  While 1\,nT = 10\,$\mu$G simply converts the field strength, many astronomical formulae are based on cgs units, which do not explicitly account for the vacuum permeability $\mu_0$.  We have made all these adjustments in the formulae and text.} $B$ in K3-50.  Nevertheless, if we assume that the magnetic field strength is related to the gas density according to the empirical formula\footnote{Although eq.\,(5) from the \citet{cru12} compilation of clouds may not necessarily apply to selections of positions within a single cloud, our intent here is merely to demonstrate how one scaling argument might work in this case.  We show in \S\ref{CF} that the estimates made here may actually be quite reasonable.}
\begin{equation} 
	B_{\rm TOT} = 49\,{\rm nT}~(n/10^{11}{\rm m}^{-3})^{0.65}
\end{equation}
from Zeeman observations of similar clouds \cite[see Fig. 6 of][]
{cru12}, and scale it to the \citet{hkp97} density model at the inner radius of the torus
\begin{equation} 
	n=5.0\times10^{11}{\rm m}^{-3}~(R/0.10\,{\rm pc})^{-1.5}  ,
\end{equation}
we obtain $B_{\rm TOT}$ = 140\,nT at the inner edge of the torus ($R$ = 0.10\,pc), dropping to 3.8\,nT at the outer edge of the cloud ($R$ = 0.75\,pc).  Combining eqs.\,(4--6),  
\begin{equation} 
	M_{\rm mag} = 270\,{\rm M}\solar~(R/0.10\,{\rm pc})^{1.025}  .
\end{equation}
Comparing this with eq.\,(2), we see that magnetic pressure at this level could provide $\sim$300\%, 133\%, or 40\% of the support required to balance gravity at 0.1\,pc, 0.22\,pc, or 0.75\,pc, respectively.  As noted above, while estimates based on eq.\,(7) do not arise from actual field strength measurements in K3-50, they do make reasonable the hypothesis that magnetic fields may contribute significant support against gravity at the inner edge of the torus.  At the same time, eq.\,(7) suggests that magnetic support is less likely to be significant at the outer edge of the cloud.

Another way to write eq.\,(4) is through the mass-to-flux ratio \citep{cn04},
\begin{equation} 
	\lambda = \frac{(M/\Phi)_{\rm obs}}{(M/\Phi)_{\rm crit}} = 0.064~\frac{N_{\rm H_2}/10^{24}{\rm m}^{-2}}{B_{\rm TOT}/{\rm nT}}  .
\end{equation}
[Here we use $\mu$\,=\,2.35 for the mean molecular weight, from a slightly more accurate value of 9\% for the He fraction by number in the gas, based on recombination-line studies of HII regions across the Galaxy \citep{s83}, instead of Crutcher et al's $\mu$\,=\,2.8, which appears to be a typographical error, for 10\% He. 
This makes the numerical coefficient in eq.\,(8) $\sim$16\% smaller than in \citet{cn04}.]

In eq.\,(8), $\lambda$$>$1 means that gravity dominates over magnetic pressure support, and the cloud is considered ``supercritical.''  Using the above estimates for $B$, $n$, and $R$ at the inner and outer edges of the molecular torus, we find that $\lambda_{\rm inner}$$\sim$0.68 while $\lambda_{\rm outer}$$\sim$140.  Again, we see that gravity may be slightly ``subcritical'' (i.e., magnetic support dominant) near the inner edge of the torus, but that near the outer edge the situation is strongly supercritical (i.e., gravity dominates).

We caution that the above estimates assume that the $B$-$n$ scaling relation in eq.\,(5) applies in the K3-50 torus.  We should compare actual magnetic field strength measurements to gauge if these estimates are realistic.  The only existing Zeeman measurements in K3-50 are for OH maser spots located near the NE corner of the UCHII region and IR cavity \citep[e.g.,][]{fraz05}, which is also near the interior NE ansa of the molecular torus, and give magnetic field strengths ranging over 300--700\,nT.  While indicating strong fields, the very restricted distribution of these maser spots limits their relevance to the thermal (whether molecular or ionised) gas.  If K3-50 becomes a future subject of a Zeeman experiment at $\sim$arcsecond or better resolution using the dense molecular tracer CN, then we would be in a better position to examine the physical roles of all these processes.  However, we explore below some direct magnetic field estimates from our polarisation measurements.

\subsubsection{The Ionised Outflow}\label{outflow}

The situation in the ionised outflow is very different, since the gas is much hotter and there is an entirely different bulk motion in the gas.  \citet{dp94} presented 14.7 GHz continuum and recombination line observations of a roughly cylindrically-shaped, bipolar ionised outflow extending 15$''$ north and south from an ultracompact HII region, and derived an outflow speed of $V$ = 35\,\kms.

Based on their data and model for the HII region, we infer a median electron density $n_e$ in the outflow area of $\sim$3$\times$10$^9$\,m$^{-3}$, although this drops quickly as one goes away from the UCHII peak, where $n_e$ $\sim$ 8$\times$10$^{11}$\,m$^{-3}$, down the $l$ = 0.63\,pc-long and 2$r$ = 0.21\,pc-wide outflow channels (measured from the source centre).  Thus, a rough estimate for the median kinetic energy density in the ionised outflow is 
\begin{equation} 
	E_{\rm kin} = 2 \left(\frac{\frac{1}{2}MV^2}{2\pi r^2l}\right) = \rho_e V^2 = 8.6\times10^{-9}\,{\rm J\,m}^{-3}  ,
\end{equation}
where we have used the median $n_e$ quoted above to calculate the equivalent mass density $\rho_e$ = $\mu\,m_H\,n_e$ and taken the mean mass per particle $\mu$ = 1.4 (now in the ionised, not molecular, gas).  This estimate for $E_{\rm kin}$ scales with $n_e$, so even though the bulk motion near the UCHII peak seems less than the average speed quoted above \citep[$\sim$10\,\kms\ from Figs.\,5 \& 6 of][]{dp94}, $E_{\rm kin}$ may be several times larger than the value in eq.\,(9) at the UCHII peak.  What {\em is} certain is that this kinetic energy density is higher closer to the outflow centre, and lower at the far ends of the outflow.


We may similarly infer a thermal energy density in the HII region, using
\begin{equation} 
	E_{\rm th} = \frac{3}{2}\rho_e\sigma^2 = n_ekT_e = 3.3\times10^{-10}\,{\rm J\,m}^{-3}  ,
\end{equation}
where $\sigma$ is the one-dimensional velocity dispersion in the HII plasma with a measured electron temperature $T_e$ = 7900\,K from \citet{dp94}.  Again, we have used our median value for $n_e$ as above as a rough estimate, but we can see that this is smaller than the kinetic energy density in the bulk outflow, by a factor of $\sim$30.  Since the ratio of energy densities $E_{\rm kin}$/$E_{\rm th}$ = $\frac{2}{3}(V/\sigma)^2$ is independent of the particle density, it is clear that the kinetic energy everywhere dominates the thermal energy in the HII region.  In other words, since the outflow is supersonic 
($M$ $\sim$ 3), it should not be surprising that the dynamical pressure exceeds the thermal pressure.

These calculations and estimates would ideally be compared to measures of the magnetic energy density in the gas (whether molecular or ionised), given by $B^2$/2$\mu_0$ (where $\mu_0$ is the permeability of free space), e.g., through Zeeman polarisation measurements of $B_{\rm los}$ at cm or mm wavelengths.  We describe below some magnetic field strength measurements for K3-50, but for now, we again estimate what magnetic field strengths {\em would be} energetically significant in the above settings.  From eq.\,(9), we find that any magnetic field in the outflow would need to rise to a level where $B^2$/2$\mu_0\sim$ 10$^{-8}$ J m$^{-3}$, or $B$ $\sim$ 150\,nT, in order to be energetically significant. 
Such a magnetic field in the {\em thermal} ionised outflow/UCHII would be unusual (i.e., discounting the several $\times$ 100 nT fields in the OH maser spots): typical measures in HII regions from (e.g.) OH Zeeman {\em absorption} observations in foreground cold molecular gas are 0.3--3\,nT \citep[e.g., 3.8\,nT in NGC 2024;][]{bc89,cru12}.

Thus, the expectation is that the kinetic energy of the ionised outflow dominates all other physical effects therein, to be considered below.

\subsubsection{Chandrasekhar-Fermi Methods}\label{CF}

Despite the lack of Zeeman magnetic field strength measurements for K3-50, we do have a rather unique set of MIR polarisation position angles ($\theta_{\rm abs}$ and $\theta_{\rm em}$), which can support some detailed analysis.

The standard Chandrasekhar-Fermi analysis \citep{cf53} connects the dispersions in polarisation angle, $\delta\theta$ = $s$ (radian), and velocity, $\delta$$V$ (cm\,s$^{-1}$), to the plane-of-sky magnetic field strength, $B_{\rm pos}$ (G).  According to \cite{cn04}, in these units the relation is
\begin{displaymath} 
	B_{\rm pos} = Q~\sqrt{4\pi\rho}~\delta V/s~;
\end{displaymath}
making the SI conversions, we have
\begin{equation}
	B_{\rm pos} = 0.85\,{\rm pT}~\sqrt{n}~\Delta V/s~,
\end{equation}
where $Q$\,=\,0.5 is a numerical factor to correct for various smoothing effects \citep[e.g., see][]{osg01}, $\rho$\,=\,$\mu\,m_{\rm H}n$ as before ($n$ in m$^{-3}$), the velocity FWHM $\Delta V$ = $\sqrt{8{\rm ln}2}~\delta V$ 
(\kms), and $s$ is now evaluated in degrees.  [Again, we use a smaller value of $\mu$ appropriate for a slightly smaller He fraction.]  If we take $\delta V\sim$ 3\,\kms\ as a median value in the \hcop\ torus from \cite{hkp97} and $n$ from eq.\,(6), then we will be able to relate $s$ in our polarisation maps to the field strength $B$.  We now examine some statistics of the dispersion before returning to this standard analysis.

One approach to evaluating the behaviour of $s$ is that of \citet{mg91}.  In their language, the goal is to identify a ``correlation length'' in the implied magnetic field orientation, within which the magnetic field directions are correlated and aligned with each other, and outside of which they are not.

Using the formalism of \citet{mg91}, we fit the distributions of polarisation position angle $\theta$ with a simple gaussian $e^{-\theta_B^2/2s^2}$ to obtain a best-fit value for the dispersion $s$ in $\theta_B$ (measured in radians).  This is a simplified version of \citet{mg91}'s analysis, since they showed that this approach gives very reliable results even for their comprehensive data (i.e., $\sim$hundreds of stellar polarisation measurements) on the Taurus molecular clouds.  We have a smaller data set of $\theta_B$, so will not need the full \citet{mg91} treatment.

\begin{figure}
\centerline{
\includegraphics[angle=0,scale=0.62]{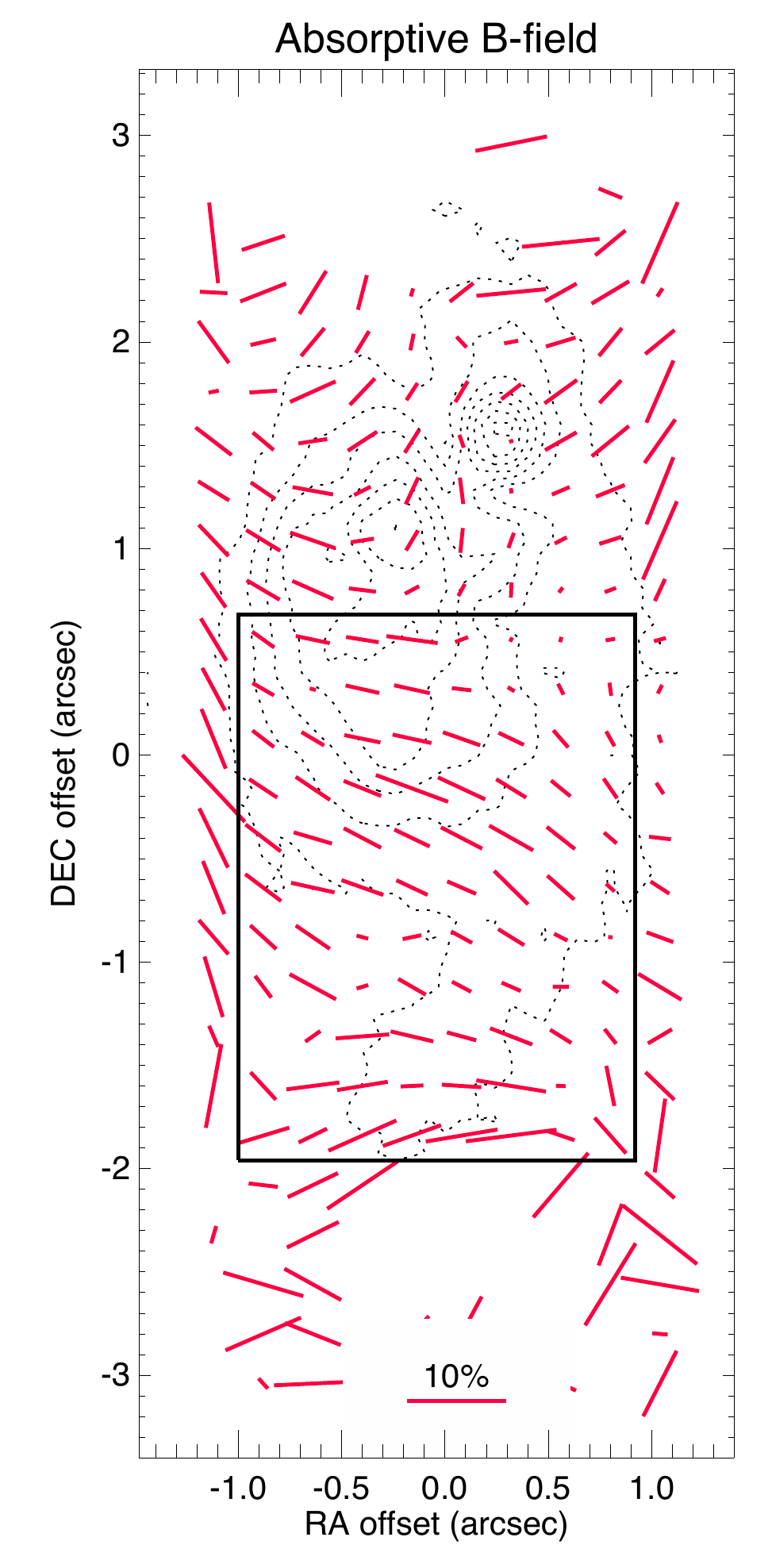}
}\vspace{-3mm}\hspace{-2mm}\centerline{
\includegraphics[angle=0,scale=0.77]{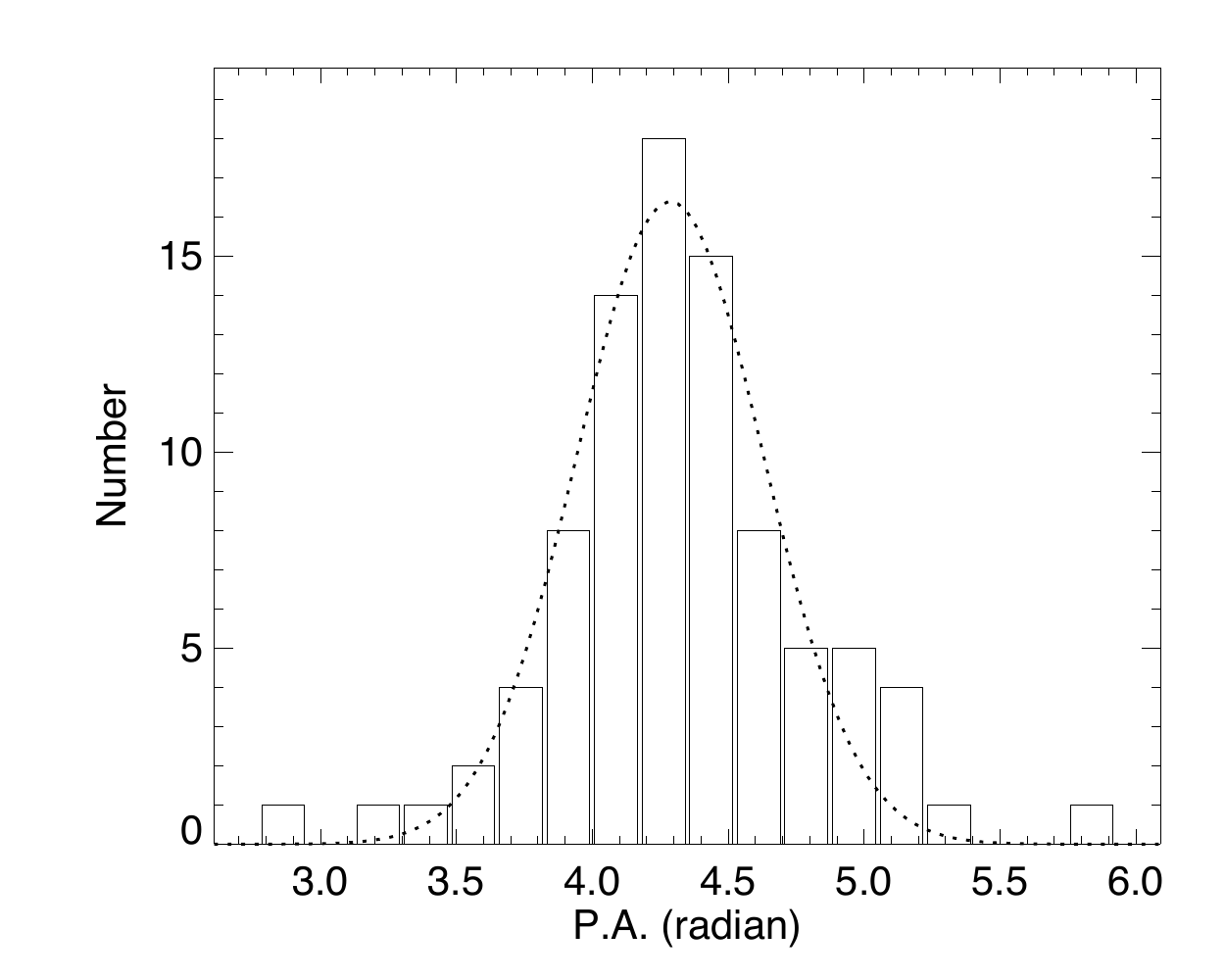}
}
\caption{(top) Embedding magnetic field orientation for $P_{\rm abs}$; these are the same red vectors indicating $\theta_{\rm B,abs}$ for the absorptive component as in Fig.\,\ref{Babsem}.  (bottom) Histogram of $\theta_{\rm B,abs}$ at all pixels within the indicated 2\farcs5$\times2''$ box in the top panel.  Such histograms can also be constructed for any smaller area within the box shown.  See text for further discussion.
\label{abshisto}}
\end{figure}

\begin{figure}
\centerline{
\includegraphics[angle=0,scale=0.62]{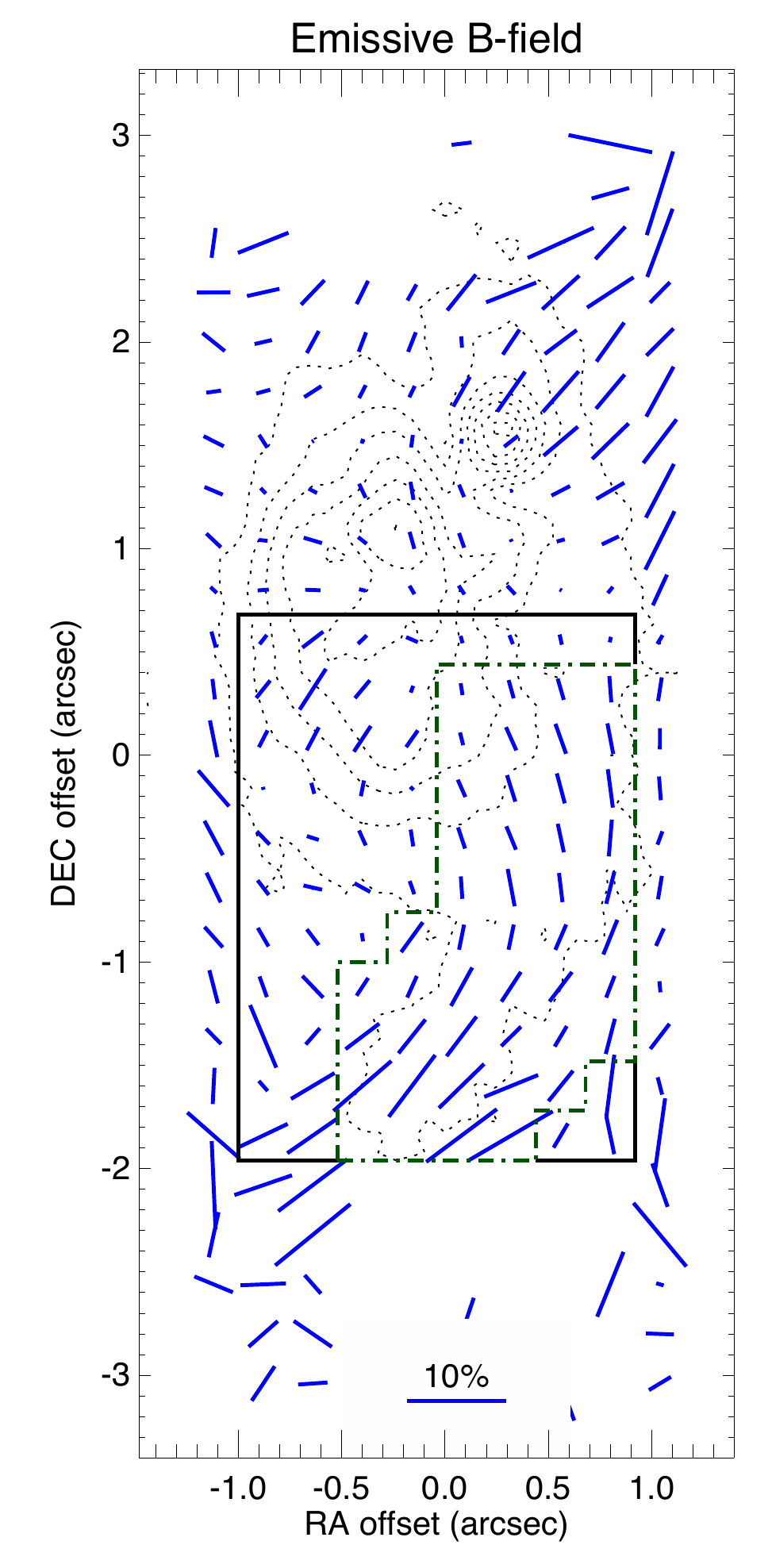}
}\vspace{-3mm}\hspace{-2mm}\centerline{
\includegraphics[angle=0,scale=0.77]{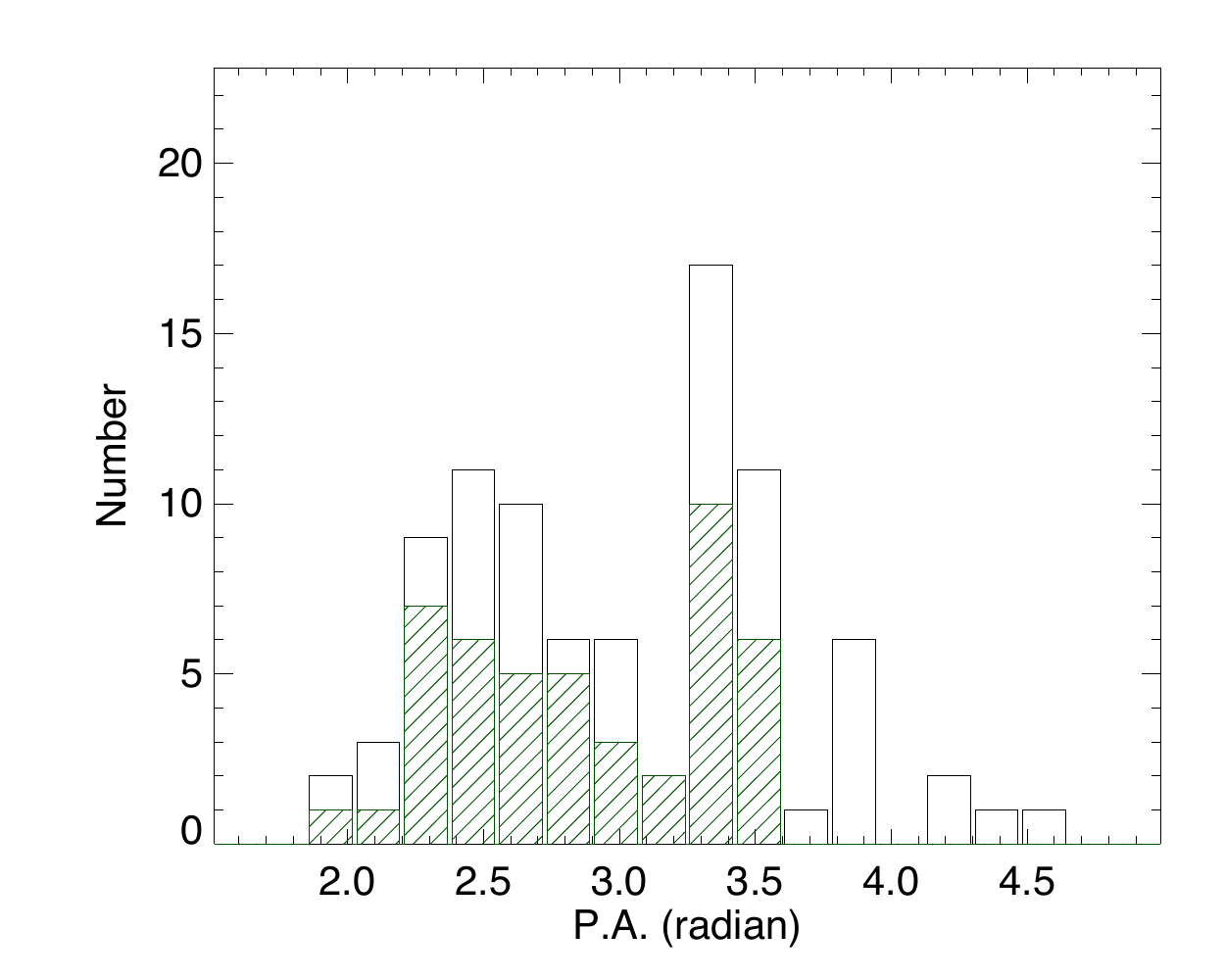}
}
\caption{Like Fig.\,\ref{abshisto}, but for the emissive component of the magnetic field.  (top) Embedding magnetic field orientation for $P_{\rm em}$; these are the same blue vectors indicating $\theta_{\rm B,em}$ for the emissive component as in Fig.\,\ref{Babsem}.  (bottom) Two overlaid histograms of $\theta_{\rm B,em}$ at all pixels within the indicated boxes in the top panel.  The black histogram is for all pixels within the rectangular black box, while the green-hatched histogram is for all pixels within the smaller, irregular green box that includes the higher-polarisation pixels on the western side of the black box.  Such histograms can also be constructed for any smaller areas within the boxes shown.  See text for further discussion.
\label{emhisto}}
\end{figure}

To evaluate $s$ in our data, we first construct histograms of the position angles from various subsets of the data as displayed in Figures \ref{Babsem}--\ref{aligned}.  The first is $\theta_{\rm abs}$ for all the vectors that can be associated with the silicate absorption in the torus as in Figures \ref{aligned}a and b, and is given in Figure \ref{abshisto}.  (The analysis for the emissive component in Figure \ref{emhisto} is slightly more complex, so we discuss it after first understanding the absorptive analysis.)  We consider only the polarisation angle distributions within a sub-window of our data (illustrated in Figs.\,\ref{abshisto} and \ref{emhisto}), where we can highlight the differences between the absorptive and emissive components.  Note that all mean quantities among these pixels that we calculate below are unweighted in our analysis, since the main sub-window that we examine here was also chosen to cover only high S/N pixels.

In the absorptive case, the distribution of $\theta_{\rm abs}$ is well-approximated by a gaussian, with a mean position angle (indicating the large-scale orientation of the magnetic field, in this case apparently aligned with the \hcop\ torus) and dispersion (which is a combination of the small measurement uncertainties, and the intrinsic degree of disorder in the large-scale magnetic field).  We first focus on the dispersion $s_{\theta_a}$ in such $\theta_{\rm abs}$ distributions.  For example, in the histogram of Figure \ref{abshisto}, the dispersion $s_{\theta_a}$ is approximately 0.35 radian.

We can also construct such histograms for subsets of this area, and then compute a dispersion in the $\theta_{\rm abs}$ distribution for each such smaller box.  The smaller these boxes are, the more choice we have of where to fit them into the large box.  Therefore, we can obtain a mean $\pm$ standard deviation in the dispersion $<$$s_{\theta_a}(A)$$>$ for all boxes of a given area $A$.  Finally, we plot these results for each box size, ranging from a minimal useful size of 3$\times$3 pixels (corresponding to 9$\times$9 pixels in unbinned maps), to the full box size shown in Figure \ref{abshisto}, obtaining the run of mean $\pm$ SD dispersions $<$$s_{\theta_a}$$>$ as a function of the chosen box size ($A^{1/2}$).

\begin{figure}
\vspace{-3mm}\hspace{-1mm}\centerline{
\includegraphics[angle=0,scale=0.76]{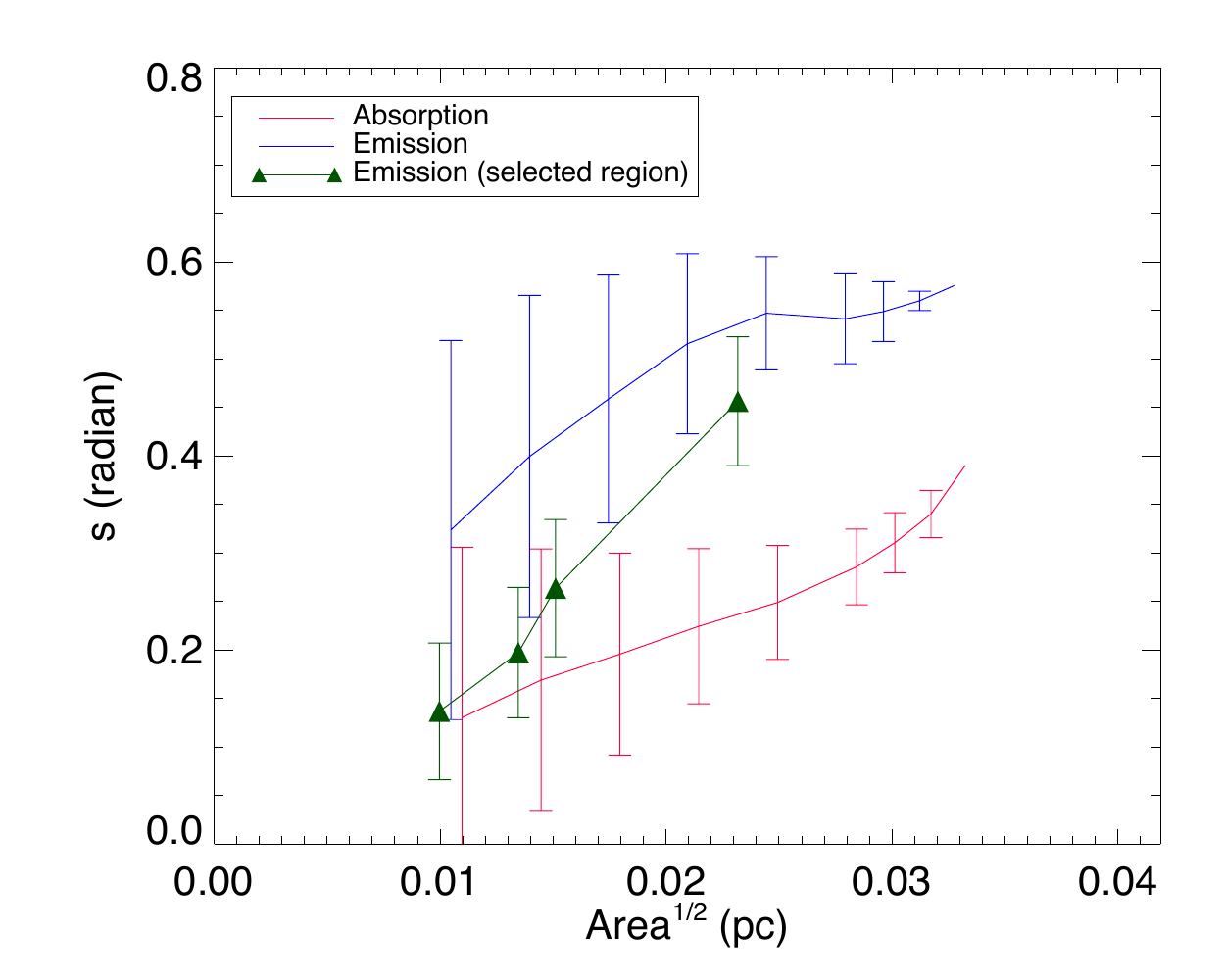}
}
\caption{Dispersion of polarisation position angles $\theta$ as a function of box size within the areas shown in Figs.\,\ref{abshisto} and \ref{emhisto}.  See text for further discussion. 
\label{sigmas}}
\end{figure}

This result for the absorptive component is plotted in red in Figure \ref{sigmas}.  Here we have statistically subtracted, in quadrature, the contribution of the $\theta$ measurement uncertainties to the measured average dispersions $<$$s_{\theta_a}(A)$$>$.  For the magnetic field associated with absorption, $s$ is quite small at all box sizes, but slowly rises with $A$.  We attribute this to the (fairly self-evident) high degree of order in the absorptive polarisation vectors in Figure \ref{aligned}b; in other words, this polarisation map shows essentially a single ``correlation length'' across the imaged area.  This makes sense in our interpretation of this magnetic field being associated with the torus, since the torus is $\sim$3$\times$ wider than our map.

We can repeat this exercise for the $\theta_{\rm em}$ distribution, except that we must first caution that the original CF method \citep{cf53} was not developed for magnetised gases in supersonic motion.  Nevertheless, numerical simulations of supersonic 
($M$ = 5--9) MHD turbulence \citep{osg01} show that it may still be reasonable to employ the CF approach to estimating magnetic field strengths in the outflow, where 
$M$ $\sim$ 3 (see \S\ref{outflow}).

In this case, the $\theta_{\rm em}$ position angles are clearly not distributed around a single mean value.  To see this, note in Figure \ref{emhisto} the distinct change in $<$$\theta_{\rm em}$$>$ between the northern and southern portions of the green box.  Note also that $\theta_{\rm em}$ for pixels outside the green box, but inside the black box, are oriented more randomly than the pixels inside the green box.  Because of this behaviour, we perform the above analysis for both samples of pixels (i.e., those from both the green and black boxes).

The black histogram in Figure \ref{emhisto} shows a much more widely distributed $\theta_{\rm em}$ due to the more variable orientations evident in the emissive polarisation map.  It turns out this is true at all box sizes within the black box, so that the plot (shown in blue) of $<$$s_{\theta_e}$$>$ vs $A^{1/2}$ in Figure \ref{sigmas} shows larger values than $<$$s_{\theta_a}$$>$ at all box sizes.  The trend with $A$ is also up, but seems to reach a plateau of $s_{\rm max}\sim0.55$ at a box size $\sim$0.025\,pc = 5,000\,AU $\sim$ 0\farcs6.

Now, when we consider the green-box pixels only, we obtain an intermediate result.  In Figure \ref{sigmas}, for smaller box sizes within the green area (plotted in green), we see relatively small $<$$s_{\theta_e}$$>$, around the same minimum value of 0.1 radian.  In other words, on small scales the magnetic field associated with the emission is about as well-ordered as the overall magnetic field associated with the absorption in the torus.  Then as the box size increases, the trend of green points rises rapidly to the same plateau in $s$ as the blue points.  Again, this is clearly due to the fact that at the larger box sizes in either the green or black areas of Figure \ref{emhisto}, we are including pixels from multiple ``correlation lengths'' of the magnetic field, so the distributions of $\theta$ represented by $s$ must get wider.

As explained by \citet{mg91}, where the CF method applies, these dispersions are related to the ratio of the disordered vs.\ ordered magnetic field strengths via
\begin{equation} 
	s = \frac{\sigma_B}{N^{1/2}B_{\perp}}   ,
\end{equation}
where $N$ is the number of magnetic field correlation lengths in the line of sight, $B_{\perp}$ is the strength of the ordered component of the magnetic field, projected into the plane of the sky (i.e., the ordered component of $B_{\rm pos}$), and $\sigma_B$ is the dispersion in the strength of the random component of $B_{\rm pos}$.  For now, we estimate $N$ from the behaviour of $s$, and so constrain somewhat the ratio $\sigma_B$/$B_{\perp}$.

From our \cc\ data, we see that typical values for $s$ lie in the range $\sim$0.1--0.6, where the lower value is clearly associated with single correlation lengths.  In those areas of our K3-50 data, we find that $\sigma_B$/$B_{\perp}$ $\sim$ 0.1--0.3, i.e., the strength of the disordered magnetic field is only 10--30\% of the strength of the ordered magnetic field.  If our interpretations are correct so far, this conclusion would therefore seem to hold for the whole of the molecular torus, and for small areas of size $\sim$ 0.01\,pc = 2,000\,AU in the ionised outflow.  This further suggests that, if the magnetic field in these areas is dynamically important, it would be the ordered component that is most likely to be relevant, rather than the disordered field.  But this may still be significant: we saw in \S\ref{torus} (e.g., by comparing eqs.\,2 and 7) that the relatively ordered magnetic field configuration in the torus implied by our data may be dynamically important in the innermost parts ($<$0.22\,pc) of this structure.  Actual CF field strength estimates in the molecular torus are discussed next.

On the other hand, the importance of magnetic fields in the ionised outflow appears less favourable.  Over larger portions of the outflow, we see evidence for at least 3 correlation lengths in the imaged area, so $\sigma_B$/$B_{\perp}$ $\sim$ 0.6$\sqrt{3}$ $\sim$ 1.  Over smaller portions of the outflow, $\sigma_B$/$B_{\perp}$ $\sim$ 0.1--0.3.  Given that any magnetic field would have to be unusually strong in the ionised gas to strongly influence the dynamics (\S\ref{outflow}), it suggests that neither a disordered nor an ordered field is likely to achieve this in the outflow, whether on smaller or larger scales.  But as we shall see, the CF analysis may give a surprising result here.

Returning now to the standard CF analysis described at the beginning of this section, we can evaluate eq.\,(11) further with these values for $s$.  Combining with eq.\,(6) we obtain for the molecular torus
\begin{equation} 
	B_{\rm pos} = (710 - 140)\,{\rm nT}~(R/0.10{\rm pc})^{-0.75}  
\end{equation} 
for areas with $s$ = 6\degree--30\degree, respectively.  What this means is that, towards the inner part ($R=0.1$\,pc) of the torus (the northern part of the box in Fig.\,\ref{abshisto}), $s\sim$ 0.3 rad (17\degree), implying that the plane-of-sky magnetic field strength $B_{\rm pos}\sim$ 250\,nT, which actually exceeds the critical value needed to resist gravity (\S\ref{torus}) by a factor of  about 1.8 ($\lambda\sim$ 0.56).  Towards the outer part of the torus ($R=0.75$\,pc), the density drops off, whence $B_{\rm pos}\sim$ 55\,nT.  This is large, but because of the greater mass contained within the larger radius, $\lambda\sim$ 15, which is still strongly supercritical (meaning gravity dominates).  These measures for $\lambda$ are very similar to the ones we estimated in \S\ref{torus}, and suggest that the general picture given there is qualitatively correct.  Therefore, if the CF analysis presented here is valid, the magnetic field at the inner edge of K3-50's molecular torus is about twice the strength suggested by the $B$-$n$ relation eq.\,(5), placing it somewhat above the mean relation in Crutcher's (2012) Figure 6.

For the ionised outflow, we can perform a similar calculation, but in this case the deduced magnetic field strength will be much smaller than in the molecular torus.  This is mainly because in eq.\,(11) the particle density is generally much lower in the outflow than the torus, and while the velocity dispersion in the flow is larger, so is the polarisation angle dispersion.  Inserting the numbers from the discussion in \S\ref{outflow} into eq.\,(11), we obtain $B_{\rm pos}\sim$ 47\,nT, which is surprisingly large given the discussion in \S\ref{outflow}, but less than the value (150\,nT) required for magnetic dominance.  However, given the inherent uncertainties involved in the use of CF methods, it is possible that in some parts of the ionised outflow (e.g., the green box in Fig.\,\ref{emhisto}), the dispersion $s$ may be low enough to indicate possible magnetic dominance in such areas.  

In summary, although there are substantial uncertainties in the application of CF methods to estimating magnetic field strengths (as embodied by the wide range of $Q$ values having been used in various studies), the weight of evidence suggests a consistent picture of K3-50: the magnetic field is probably strong enough to support the cloud against gravity in the inner parts of the molecular torus ($R<0.2$\,pc), but is not strong enough to resist gravity outside this area.  While the magnetic field in the outflow also seems somewhat too weak to dominate the overall energetics, it may nevertheless be important in some locations.

\subsection{Comparison with Models}

Theoretical simulations of star formation have been a vigorous area of activity for many years.  Recent examples show that, in order to understand the magnetic field configurations inferred from polarisation measurements of massive protostellar objects, it may be necessary to model fully three-dimensional geometries that include misaligned rotation and magnetic axes viewed from arbitrary perspectives \citep[e.g.,][]{mmht06,kmt12,s12}.  Our results for K3-50 can be compared with such models on the physical scales represented in our maps, namely $\sim$1,000--20,000\,AU.

That the magnetic field associated with the absorptive component of the 10$\mu$m silicate feature seems to be well-aligned with the toroidal geometry of \citet{hkp97}'s molecular disk, and that the magnetic field strength in the inner 0.2\,pc of the torus may be dynamically significant, 
suggests that the magnetic field may have played an important role in the evolution of this star-forming cloud, or may even have dominated the formation of these structures.

On the other hand, \citet{mmht06} showed that one could reproduce such features, especially a toroidally-aligned magnetic field on scales of $\sim$2000--20,000\,AU, only if the cloud is rotation-dominated and the magnetic field relatively weak (their ``C'' models).  For the strong-field case, their ``D'' and ``F'' models show that toroidal alignment only occurs on much smaller scales, \lapp300\,AU.  In support of this weak-field scenario, our estimates in \S\ref{torus} show that, outside the inner edge of the torus modelled by \citet{hkp97} (i.e., on the larger scales that may be more closely related to the ``initial conditions'' for this star-forming cloud), the magnetic field is indeed unlikely to be dominant.


The review of \citet{w07} showed that, on the larger (pc) scale, magnetic fields are preferentially aligned toroidally in (pseudo-)disks, but poloidally along outflow axes, and we see here that K3-50 provides a classic example of such alignments.  It is tempting to go further and attribute these alignments in K3-50 on the larger scale ($>$1\,pc) to the general interstellar magnetic field setting the orientation of gas collapse along field lines to form the molecular torus; also on the smaller scale ($<$1\,mpc), to the magnetically-driven outflow mechanism of \citet{bp07} and similar models of that ilk.  The justification is that, although our \cc\ K3-50 data do not directly probe either of these scales, physical laws must still be ``obeyed'' in connecting these scales through the 0.01--0.1\,pc range revealed here.

However, there are some difficulties in making this connection.  For example, optical polarisation data of stars in the direction of K3-50 over the relevant distance range \citep{h00} suggest that the larger-scale magnetic field in the vicinity of K3-50 is not so conveniently oriented.  Also, as described above, the formation of the dense clumps responsible for cluster formation may be magnetically supercritical, at least on the larger scales.  Likewise, many of the protostellar outflow models tend to be for the lower-mass/single star case, and also are typically computed over 10$^{2-3}$ yr timescales, rather than for a massive cluster over the 10$^{4-5}$ yr timescale of the outflow or disk kinematics (\S\S\ref{torus}--\ref{outflow}).  Finally, in the ionised outflow, it seems somewhat surprising (given past Zeeman measurements in HII regions) that the magnetic field strength may be strong enough in some locations to influence the highly supersonic motions in this volume (\S\ref{outflow}).  On the other hand, it is certainly clear that the magnetic field here is not ``primordial'' to the cloud's formation, and must be dynamically relevant to the outflow itself.  This suggests that other magnetic field measurements in HII regions must be attributed with care to the appropriate dynamical setting.

Thus, challenges in understanding these data and interpretations remain.  As we have seen here, new instruments like \cc\ will certainly be pivotal in improving this understanding.

\section{Conclusions}

Decomposition of magnetic field components using mid-IR imaging polarimetry is a very clever ``old'' technique, made new and more powerful with \cc.  By carefully considering the polarisation information from this new instrument, we can conduct pioneering science such as ``polarisation tomography'' of magnetic fields to decipher their three-dimensional geometry in sources, as well as constrain magnetic field strengths in these same areas.

We have presented new imaging polarimetry results using the mid-infrared silicate feature for the luminous UCHII region K3-50, which give new constraints on the magnetic field structure, strength, and role in this massive star-cluster forming cloud.  Our main results are:
\begin{enumerate}
\item[1.] We find that the respective magnetic field components are substantially aligned with both the $\sim$0.4-pc scale molecular torus that surrounds the young star cluster, and with the ionised outflow that is being driven from the same proto-cluster.
\item[2.] Scaling arguments and comparison with models suggest that the magnetic field may be energetically important in the inner parts of the molecular torus, but unimportant outside a radius $\sim$0.25\,pc, where the torus is instead probably rotationally-supported against gravity.
\item[3.] In the outflow, the magnetic field is probably not globally dominant, but it may be important in some locations, such as in confining the outflow along the cavity walls.
\end{enumerate}
Future field-strength measurements, such as from Zeeman experiments with OH or CN lines, would be very helpful to confirm these results.

Similar observations of other sources with \cc\ will allow us to examine what role magnetic fields play in star formation, stellar and protoplanetary disks, AGN, and other astrophysical situations, providing new constraints on theoretical models for such objects.

\section{Acknowledgments}

We are grateful to the GTC staff for their outstanding support of the commissioning and science operations of \cc.  The work presented in this paper has been supported in part by NSF awards AST-0903672 and AST-0908624 to CMT.  PJB acknowledges support from NASA/JPL contract RSA-1464327, NSF grant AST-1312597, and the UF Astronomy Department.  CMW acknowledges support from Australian Research Council Future Fellowship FT100100495.  This research is based on observations using \cc\ at the Gran Telescopio Canarias, a partnership of Spain, Mexico, and the University of Florida, and located at the Spanish Observatorio del Roque de los Muchachos of the Instituto de Astrof\'isica de Canarias, on the island of La Palma.  We warmly thank the anonymous referee for a very positive review, and helpful comments which improved the presentation of this paper.

Facilities: 
{GTC(CanariCam)}.



\end{document}